\documentstyle[12pt]{article}
\let\ssection=\section
\renewcommand{\section}{\setcounter{equation}{0}\ssection}

\newcommand{\be}{\begin{enumerate}}
\newcommand{\ee}{\end{enumerate}}
\newcommand{\bi}{\begin{itemize}}
\newcommand{\ei}{\end{itemize}}
\newcommand{\beq}{\begin{equation}}
\newcommand{\eeq}{\end{equation}}
\newcommand{\beqa}{\begin{eqnarray}}
\newcommand{\eeqa}{\end{eqnarray}}

\newcommand{\eq}[1]{(\ref{#1})}
\newcommand{\eqs}[2]{(\ref{#1}--\ref{#2})}
\newcommand{\ie}{{\em i.e.,\ }}

\renewcommand{\a}{\alpha}                   
\renewcommand{\b}{\beta}                    
\renewcommand{\d}{\delta}		

\newcommand{\g}{\gamma}

\newcommand{\s}{\sigma}

\newcommand{\bra}[1]{\langle #1 |}
\newcommand{\braket}[2]{\bra #1 #2\rangle}
\newcommand{\ket}[1]{| #1 \rangle}
\newcommand{\ketbra}[2]{\ket #1\bra #2}
\newcommand{\op}[1]{\widehat{#1}}
\newcommand{\tr}{{\rm tr}}

\newcommand\mathC{\mkern1mu\raise2.2pt\hbox{$\scriptscriptstyle|$}
                {\mkern-7mu\rm C}}
\newcommand{\mathR}{{\rm I\! R}}                

\newcommand{\C}{{\tilde{C}}}
\newcommand{\D}{{\cal D}}
\newcommand{\U}{{\cal U}}
\newcommand{\UP}{{\cal UP}}
\renewcommand{\S}{{\cal S}}
\renewcommand{\H}{{\cal H}}
\renewcommand{\L}{{\cal L}}

\newcommand{\h}[1]{({#1}_{t_1},{#1}_{t_2},\ldots,{#1}_{t_n})}

\begin{document}

\begin{titlepage}
\hspace{8truecm} Imperial/TP/93--94/35

\hspace{8truecm} DAMTP 94--35

\begin{center}
        {\large\bf Quantum Temporal Logic and Decoherence Functionals in
the\\[0.2cm]
                   Histories Approach to Generalised Quantum Theory}
\end{center}
\vspace{1 truecm}
\begin{center}
        C.J.~Isham\footnote{Permanent address: Blackett
        Laboratory, Imperial College, South Kensington, London SW7
        2BZ; email: c.isham@ic.ac.uk}\\[0.4cm]
        Isaac Newton Institute for Mathematical Sciences\\
        University of Cambridge\\
        20 Clarkson Road\\
        Cambridge CB3 0EH\\
        United Kingdom\\
\medskip
\begin{center} and \end{center}
\medskip
        N.~Linden\footnote{email: n.linden@newton.cam.ac.uk}\\[0.4cm]
        D.A.M.T.P.\\
        University of Cambridge\\
        Cambridge CB3 9EW\\
        United Kingdom\\
\end{center}

\begin{center} April 1994\end{center}

\begin{abstract}
We analyse and develop the recent suggestion that a temporal form of
quantum logic provides the natural mathematical framework within which
to discuss the proposal by Gell-Mann and Hartle for a generalised form
of quantum theory based on the ideas of histories and decoherence
functionals. Particular stress is placed on properties of the space of
decoherence functionals, including one way in which certain global
and topological properties of a classical system are reflected in a
quantum history theory.
\end{abstract}

\end{titlepage}

\section{Introduction}
 The main aim of this paper is to explore the algebraic structure
underlying the generalised quantum scheme proposed by Gell-Mann and
Hartle \cite{GH90a,GH90b,GH90c,Har91a,Har91b,GH92,Har93a}. This scheme
is a natural development of the decoherent histories approach to
standard quantum theory put forward by Griffiths \cite{Gri84} and by
Omn\`es \cite{Omn88a,Omn88b,Omn88c,Omn89,Omn90,Omn92}. We shall
describe how the algebraic structure underlying this scheme parallels
that of quantum logic \cite{Ish94a}, with particular emphasis on the
space of decoherence functionals viewed as the analogue of the space
of states in standard quantum theory.

    A major motivation for studying schemes of this type in the
context of quantum gravity is the well-known `problem of time'.  In
particular, the unsolved question of whether the normal concept of
time is fundamental or one that emerges only above the Planck
scale in some coarse-grained way. If the latter is true it seems
reasonable to suggest that the usual notion of a continuum space
manifold may also be applicable only in some semi-classical sense: a
view that is difficult to reconcile with the standard canonical
approach to quantum gravity which employs a fixed background
three-manifold.

    This difficulty reinforces the idea of developing a more
spacetime-oriented approach to quantum gravity (what is sometimes
called the `covariant', as opposed to `canonical', perspective
\cite{DeW67a,DeW67b}) albeit one in which the classical ideas
of space and time play no fundamental role. A non-standard quantum
scheme of this sort could permit the introduction of very generalised
spacetime concepts like quantum topology, causal sets \cite{Sor91b},
and various other discrete models. A key supposition is that the
familiar spacetime concepts `emerge' well above the Planck scale, as
must also the Hilbert space mathematical formalism of normal quantum
theory which is tied so closely to the standard picture of space and
time. This suggests that the infamous Copenhagen interpretation might
also apply only in some approximate sense. Indeed, a major driving
force behind the original studies of decohering histories was a desire
to find an interpretation of quantum theory with a philosophical
flavour that is more realist than is the usual one. Such a move from
observables to `beables' \cite{Bel87} is particularly attractive in
any theory that aspires to address issues in quantum cosmology.

    In this paper we focus on the approach to generalised quantum
theory advocated by Gell-Mann and Hartle in which the notion of a
`history' may not just be a time-ordered string of events or
propositions: rather, it can appear as a fundamental theoretical
entity in its own right. For example, \cite{Har93a}  a Lorentzian
manifold $({\cal M},g)$ that is not globally-hyperbolic cannot be
regarded as a time-ordered sequence of three-geometries as there is no
global time function with which to construct an appropriate
one-parameter family of foliations of $\cal M$. In this sense, the
Lorentzian metric $g$ is not a `history' but, nevertheless, $g$ may
be perfectly acceptable as a geometry of space-time, and hence
as a `generalised history' of the universe.

    This example illustrates the main point well: in classical or
quantum gravity a `history' is simply a potential configuration for
the universe as a whole (or, perhaps more precisely, a proposition
about such configurations), including any quasi space-time structure
it may possess. Thus a `history' might involve a causal set or other
discrete structure, or it could just be a simple continuum manifold
with associated geometric structure.

    Our approach to the Gell-Mann and Hartle (hereafter abbreviated to
GH) scheme is based on \cite{Ish94a} whose central idea is to use
`quasi-temporal' quantum logic to provide a firm mathematical footing
for the scheme. Indeed, as we shall see, by rewriting the GH rules in
a certain way, the potential connection with quantum logic becomes
very clear.

    The programme of standard quantum logic began with a seminal paper by
von Neumann and Birkhoff \cite{VNB36}, and has been developed by many
authors since (an excellent review is \cite{BC81}). The starting point
was the claim by von Neumann and Birkhoff that the essential
differences between a classical system and a quantum system are captured
by the structures of their associated spaces of propositions about the
system at a fixed time: in classical physics this is the Boolean
algebra of measureable subsets of the classical phase space; in the
quantum case it is the non-distributive lattice of projection
operators on the Hilbert space $\H$ or, equivalently, the lattice of
closed subsets of $\H$.

    In the latter context we recall that the following lattice
operations are defined (where $\H_P\subset\H$ and $\H_R\subset\H$
denote the ranges of projection operators $P$ and $R$ respectively):
(i) the partial ordering operation $P\leq R$ if and only if
$\H_P\subseteq\H_R$; (ii) the complementation (\ie `not') operation
$\neg P:=1-P$ (so that $\H_{\neg P}$ is $\H_P^\perp$); (iii) the meet
$P\wedge R$ is defined as the projector onto $\H_P\bigcap\H_R$; and
(iv) the join $P\vee R$ is defined as the projector onto the closure
of the linear span of $\H_P\bigcup\H_R$. This structure is interpreted
physically by saying that if the state of the system is the density
matrix $\rho$ and if a measurement is made of the proposition
represented by a projection operator $P$, then the probability that the
proposition will be found to be true is ${\rm tr}(P\rho)$.

    One aim of the quantum logic programme was to explore the
possibility of associating the propositions of a quantum system with a
lattice $\L$ that is not just the projection lattice ${\cal P}(\H)$ of
a Hilbert space $\H$. Thus, dual to such a lattice $\L$, there is a
space $\S$ of {\em states\/} where $\s\in\S$ is defined to be a
real-valued function on $\L$ with the properties
 \be
    \item {\em Positivity\/}: $\s(P)\ge0$ for all $P\in\L$ and
            $\s\in\S$.
    \item {\em Additivity\/}: if $P$ and $R$ are disjoint (defined to
          mean that $P\leq\neg R$) then, for all $\s\in\S$,
          $\s(P\vee R)=\s(P)+\s(R)$. This requirement is usually extended
          to include countable collections of propositions.
    \item {\em Normalisation\/}: $\s(1)=1$
\ee
 where $1$ is the unit proposition for which $\H_1=\H$ (there is also
a null proposition $0$ with $\H_0=\{0\}$). Physically, $\s(P)$ is
interpreted as the probability that proposition $P$ is true (or, more
operationally, will be found to be true if an appropriate measurement
is made) in the state $\s$. The trivial propositions $0$ and $1$ are
respectively false and true with probability one in any state.

    For the purposes of our later discussion we note that there
has been considerable debate about the physical meaning
of the $\vee$ and $\wedge$ operations when applied to propositions
that are not compatible. This has lead a number of authors to suggest
weakening the idea of a lattice so that, for example, the join
operation is defined only on pairs of propositions $P$ and $R$ that
are {\em disjoint\/}, in which case it is customary to write the join
as $P\oplus R$ rather than $P\vee R$. Thus, in schemes of this type,
one has only (i) a partial ordering operation; (ii) an
orthocomplementation operation $\neg$; and (iii) a notion that certain
pairs $P,R\in\L$ are {\em disjoint\/}, written $P\perp R$, and that
such pairs can be combined to give a new proposition $P\oplus R$
interpreted as `$P$ {\em or\/} $R$'. The minimal useful structure of this
type appears to be an {\em orthoalgebra\/} \cite{FGR92} in which the
partial-ordering and disjoint-join operations are related by the
condition $P\leq R$ if and only if there exists $V\in\L$ such that
$R=P\oplus V$.

    A slightly stronger concept is of an {\em orthomodular poset\/}, which
is essentially an orthoalgebra with the extra requirement that $P\oplus
R$ is the unique least upper bound for $P$ and $R$.

    In standard quantum logic, time evolution appears as a
one-parameter family of automorphisms of the lattice of propositions;
in a Hilbert space this is implemented by the familiar family
$t\mapsto U(t):=e^{-iHt/\hbar}$ of unitary operators. However, our
scheme is quite different since we propose to use an orthoalgebra to
model the space of propositions about {\em histories\/} of the system;
a proposal that suggests the use of a quantum version of {\em
temporal\/} logic, rather than the single-time logic employed in
standard quantum theory.

    Temporal logic has not featured strongly in existing studies of
quantum logic with the notable exception of the work of Mittelstaedt
and Stachow  \cite{Mit77,Mit78,Sta80,Sta81} who developed a theory of
the logic of sequential propositions using sequential conjunctions of
the type ``$A$ is true {\em and then\/} $B$ is true {\em and then\/}
$\ldots$'', plus other temporal connectives such as `or then' and
`sequential implication'. The approach adopted here and in
\cite{Ish94a} is partly an extension of this work to include
`quasi-temporal' situations in which the notion of time-evolution is
replaced by something that is much broader and which, for example,
could include space-time causal relations as well as ideas of quantum
topology. Ultimately we are interested in situations where there is no
prior temporal structure at all. However, in all cases, a key
ingredient in our approach is the observation that the statement that
a certain universe (\ie history) is `realised' is itself a
proposition, and therefore the set $\UP$ of all such
history-propositions might possess the structure of an orthoalgebra or
lattice that is analogous to the lattice of single-time propositions
in standard quantum logic. In particular, a history proposition might
be representable by a {\em projection\/} operator in some Hilbert
space. But, whether this is true or not, the heart of our approach is
a claim that, if $\D$ denotes the set of decoherence functionals (a
decoherence functional is a complex-valued function of pairs of
histories that measures their mutual quantum interference), the pair
$(\UP,\D)$ plays a role in history theory that is analogous to that of
the pair $(\L,\S)$ in normal, fixed-time quantum theory.

    The plan of the paper is as follows. We begin with a version of
the Gell-Mann and Hartle rules for generalised quantum theory that is
based on the ideas in \cite{Ish94a} but presented in a way that is
intended to make the use of quantum-logical techniques particularly
natural. Then, using a simple two-time model, we show how the
history-propositions in standard quantum theory can indeed be
represented by projection operators on a new Hilbert space, and that
this throws useful light on some of the manipulations performed by
Gell-Mann and Hartle in their development of the scheme. We also
review the way in which `quasi-temporal' structure can be incorporated
as a generalisation of the partial semigroup that is associated with
the standard time axis.

    This is followed by a discussion of several properties of
decoherence functionals in standard quantum theory with a view to
seeing what can, or should, be incorporated into the general scheme. We
consider several ways in which decoherence functionals can be
constructed, and we show that several natural-sounding properties that
might be posited in the general theory are in fact violated even in
standard quantum theory. This is followed by a discussion of how
certain global or topological properties of the classical system are
reflected in a quantum history theory. A key issue which we address is
how the familiar $\pi_1(Q)$ effects in path-integrals over a
configuration space $Q$ are manifested in the more abstract histories
approach to the quantum theory. This framework throws a new light on
the origin of such effects and how they relate to general global
properties of the classical system.

    Finally, a word about notation. It is sometimes useful to
distinguish between a proposition $P$ and the projection operator that
represents it, in which case the latter will be written as $\op P$. But
in general `hats' will be avoided for the sake of typographical
clarity.

\section{Generalised Quantum Theory}
\subsection{Histories in standard quantum theory}
 In developing their abstract notion of a `history', Gell-Mann and
Hartle were guided partly by ideas that arise naturally in
path-integral quantisation using paths in a configuration space $Q$,
and partly by the idea of a history as a time-ordered sequence of
projectors in a Hilbert-space based quantum theory. Both approaches
suggest certain logical-type operations in the space of all histories.

    Let us begin by reviewing briefly how histories are handled in a
standard quantum theory defined on a Hilbert space $\H$. As usual,
observable quantities are represented by self-adjoint operators on
$\H$ and, to avoid ambiguities associated with overall unitary
transformations, we will suppose that (i) a fixed labelling of
operators with specific physical quantities has been established; and
(ii) the Schr\"odinger representation is used throughout unless the
contrary is indicated explicitly. A proposition concerning the values
of an observable is represented mathematically by a projection
operator $P$ on $\H$. Note that, by virtue of our  assumptions, the
representation of a particular proposition by a specific projector
does not depend on time: the time dependence is carried solely by the
evolving state.

    Suppose now we are concerned with a sequence of propositions
$\a:=\h{\a}$ that refer to the results of measurements made at times
$t_1,t_2,\ldots,t_n$ where $t_1<t_2<\cdots<t_n$, and which is such
that each proposition is non-trivial, {\rm i.e.,} not equal to $0$ or
$1$. We shall refer to this set as a potential {\em homogeneous
history\/} of the system with the understanding that the history is
`realised' if  each proposition $\a_{t_i}$ is found to be true at the
corresponding time $t_i$. Thus, in the language of temporal logic,
$\h{\a}$ is the {\em sequential conjunction\/} ``$\a_{t_1}$ is true at
time $t_1$ {\em and then} $\a_{t_2}$ is true at time $t_2$ {\em and
then} $\ldots$ {\em and then} $\a_{t_n}$ is true at time $t_n$''.

    It is convenient to think of a homogeneous history $\a$ as a
proposition-valued function $t\mapsto \a_t$ on the time-line $\mathR$
which is equal to the unit proposition $1$ for all but a finite number
of time values. We shall call this set $\{t_1,t_2,\ldots,t_n\}$ at
which the history is `active' the {\em temporal support\/} of the
homogeneous history. The situation can arise in which the result of
some logical operation on homogeneous histories is a history in which
at least one of the propositions in the sequence is the null
proposition $0$. All such histories are defined to be equivalent to
each other and to the null history.  Note that the subscript $t_i$
serves merely to label the different propositions and to remind us of
the times at which they are asserted: the associated projection
operators $\op\a_{t_i}$ are in the {\em Schr\"odinger\/}
representation.

    A standard result in conventional quantum theory using
state-vector reduction is that if the state at some initial time
$t_0$ is the density matrix $\rho$, then the joint probability of
finding all the associated properties in an appropriate sequence of
measurements is
 \beq {\rm Prob}(\a_{t_1},\a_{t_2},\ldots,\a_{t_n};\rho_{t_0})=
              \tr(\C_\a^\dagger\rho \C_\a)                \label{Prob:a1-an}
\eeq
where the `class' operator $\C_\a$ is defined by
\beq
    \C_\a:=U(t_0,t_1)\a_{t_1} U(t_1,t_2)\a_{t_2}\ldots U(t_{n-1},t_n)
        \a_{t_n}U(t_n,t_0)                \label{Def:C_a}
\eeq
 and where $U(t,t')=e^{-i(t-t')H/\hbar}$ is the unitary
time-evolution operator from time $t$ to $t'$; in particular, $U$
satisfies the evolution law $U(t_1,t_2) U(t_2,t_3)= U(t_1,t_3)$ for
all $t_1$, $t_2$ and $t_3$. Note that, to ease the later discussion
of the relation with temporal logic, our operator $\C_\a$ is the
adjoint of the operator $C_\a$ used by Gell-Mann and Hartle. We note
also in passing that $\C_\a$ is often written as the product of
projection operators
 \beq
     \C_\a=\a_{t_1}(t_1)\a_{t_2}(t_2)\ldots\a_{t_n}(t_n)
\eeq
where $\a_{t_i}(t_i):= U(t_0,t_i)\a_{t_i} U(t_0,t_i)^\dagger$ is
the Heisenberg picture operator defined with respect to the fiducial
time $t_0$.

    The main assumption of the consistent-histories interpretation of
quantum theory is that, under appropriate conditions, the probability
assignment \eq{Prob:a1-an} is still meaningful for a {\em closed\/}
system, with no external observers or associated measurement-induced
state-vector reductions (thus signalling a move from `observables' to
`beables'). The satisfaction or otherwise of these conditions (the
`consistency' of a complete set of histories: see below) is determined
by the behaviour of the {\em decoherence functional\/}
$d_{(H,\rho)}$. This is the complex-valued function of pairs of
homogeneous histories $\a=\h{\a}$ and
$\b=(\b_{t_1'},\b_{t_2'},\ldots,\b_{t_m'})$ defined as
 \beq
    d_{(H,\rho)}(\a,\b)=\tr(\C_\a^\dagger\rho \C_\b)         \label{Def:d}
 \eeq
where the temporal supports of $\a$ and $\b$ need not be the same. Note
that, as suggested by the notation $d_{(H,\rho)}$, both the initial
state and the dynamical structure (\ie the Hamiltonian $H$) are coded
in the decoherence functional. In our approach, a history $\h{\a}$
itself is just a `passive', time-ordered sequence of propositions.

    Two important concepts in the history formalism are {\em
coarse-graining\/} and {\em disjointness\/}. It will be helpful at
this point to give the relevant definitions in a way that emphasises
their relation to logical operations.
 \bi
 \item  A homogeneous history  $\b:=\h{\b}$ is {\em coarser\/} than
another such $\a:=(\a_{t_1'},\a_{t_2'},\ldots,\a_{t_m'})$ if the
temporal support of $\b$ is equal to, or a proper subset of, the
temporal support of $\a$, and if for every $t_i$ in the support of
$\b$, $\a_{t_i}\leq\b_{t_i}$ where $\leq$ denotes the usual ordering
operation on projection operators. The ensuing partial ordering on the
set of homogeneous histories is written as $\a\leq\b$.

 \item Two homogeneous histories  $\a:=(\a_{t_1'},\a_{t_2'},
\ldots,\a_{t_m'})$ and $\b:=\h{\b}$ are {\em disjoint\/} if their
temporal supports have at least one point in common, and if for at
least one such point $t_i$ the proposition $\b_{t_i}$ is disjoint
from $\a_{t_i}$, \ie the ranges of the two associated projection
operators are orthogonal subspaces of $\H$ so that
$\a_{t_i}\b_{t_i}=0=\b_{t_i}\a_{t_i}$.

 \item The {\em unit\/} history assigns the unit proposition to every
time $t$. The {\em null\/} history is represented by any history in
which at least one of the propositions is the zero proposition (all such
null histories are regarded as equivalent). It follows that
$0\leq\a\leq 1$ for all homogeneous histories $\a$.
 \ei

    A central role in the GH-scheme is played by the possibility of
taking the join $\a\oplus\b$ of a pair $\a$, $\b$ of disjoint
histories with the heuristic idea that $\a\oplus\b$ is `realised' if
and only if either $\a$ is realised or $\b$ is realised. This
operation is straightforward if $\a$ and $\b$ have the same temporal
support and differ in their values at a single time point $t_i$ only.
In this case, if  $\a_{t_i}$ and $\b_{t_i}$ are disjoint projectors
then $\a\oplus\b$ is equal to $\a$ (and therefore $\b$) at all time
points except $t_i$, where it is $\a_{t_i}+\b_{t_i}$. Thus
$\a\oplus\b$ is a homogeneous history and satisfies the relation
$\C_{\a\oplus\b} =  \C_\a+\C_\b$.

    However, it is important to be able to define $\a\oplus\b$ for
homogeneous histories $\a,\b$ that are disjoint in a less restrictive
way. Similarly, one would like to have available a negation operation
$\neg$ so that, heuristically, $\neg\a$ is realised if and only if
$\a$ is definitely not realised. But both operations take us outside
the class of homogeneous histories.

    Gell-Mann and Hartle approached this issue by considering the case
of a path-integral on a configuration space $Q$. The
analogue of the decoherence functional \eq{Def:d} is
 \beq
    d(\a,\b):=\int_{q\in\a,q'\in\b}Dq\,Dq'\,e^{-i(S[q]-S[q'])/\hbar}
            \d\big(q(t_1),q'(t_1)\big)\rho\big((q(t_0),q'(t_0)\big)
                            \label{Def:dPI}
 \eeq
where the integral is over paths that start at time $t_0$ and end at
time $t_1$, and where $\a$ and $\b$ are subsets of paths in $Q$. In
this case, to say that a pair of histories $\a$ and $\b$ is disjoint
means simply that they are disjoint subsets of the path space of $Q$,
in which case $d$ clearly possesses the additivity property
 \beq
        d(\a\oplus\b,\g)=d(\a,\g)+d(\b,\g)
 \eeq
for all subsets $\g$ of the path space. Similarly, $\neg\a$ is
represented by the complement of the subset $\a$ of path space, in
which case the decoherence functional satisfies
 \beq
        d(\neg\a,\g)= d(1,\g)-d(\a,\g)
 \eeq
where $1$ denotes the entire path space (the `unit' history).

    Gell-Mann and Hartle noted that these properties can be replicated
for the decoherence functionals on strings of projectors
by postulating that, when computing a decoherence functional,
one should use the class operators $\C_{\a\oplus\b}$ and $
\C_{\neg\a}$ that are essentially {\em defined\/} by
 \beq
         \C_{\a\oplus\b}:= \C_\a+ \C_\b         \label{Def:Ca+b}
 \eeq
and
 \beq
         \C_{\neg\a}:=1-\C_\a                   \label{Def:Cnega}
 \eeq
 This procedure sounds reasonable but in some respects it is rather
curious. For example, the right hand side of \eq{Def:Ca+b} is a
concrete operator that `represents' $\a\oplus\b$ as far as calculating
decoherence functionals is concerned, but it is not clear what
it is that is actually being represented! A homogeneous history
$\a$ is a time-ordered sequence $\h{\a}$ of propositions $\a_{t_i}$
but there is no immediate analogue for the `inhomogeneous' history
$\a\oplus\b$. A similar remark applies to the negation $\neg\a$ of a
homogeneous history $\a$.

    On the other hand, \eqs{Def:Ca+b}{Def:Cnega} look familiar if one
considers the standard operations in normal single-time quantum logic.
Thus if the projection operators $\op P$ and $\op R$ represent
single-time propositions $P$ and $R$, and if $\op P$  and $\op R$ are
disjoint, then the proposition $P\oplus R$ is represented by $\op
P+\op R$. Similarly, the proposition $\neg P$ is represented by the
projection operator $1-\op P$. Thus \eq{Def:Ca+b} and \eq{Def:Cnega}
do suggest that {\em some\/} sort of logical operation is involved,
but they cannot simply be justified by invoking the standard
operations on projectors since, generally speaking, a product of
projection operators like $\C_\a$ is not itself a projector.

    On reflection, these observations suggest that what is needed is a
quantum form of {\em temporal\/} logic and, as we shall show later,
this is indeed the case. However, less us first present the GH-rules
for generalised quantum theory in a way that demonstrate their natural
connection with the types of algebraic structure used in conventional
quantum logic \cite{Ish94a}.

\subsection{A minimal version of the Gell-Mann and Hartle scheme}
\label{SSec:MinGH}
 The Gell-Mann and Hartle `axioms' \cite{Har93a} constitute a new
approach to quantum theory in which the notion of history is ascribed
a fundamental role; \ie a `history' in this generalised sense (which
is intended to include analogues of both homogeneous and inhomogeneous
histories) is an irreducible structural entity in its own right that
is not necessarily to be derived from time-ordered strings of
single-time propositions. We shall present our version of these axioms
in a way that brings out the relation to quantum logic.

\be
 \item The fundamental ingredients in the theory are (i) a space $\UP$
of propositions about possible `histories' (or `universes'); and (ii)
a space $\D$ of {\em decoherence functionals\/}. A
decoherence-functional is a complex-valued function of pairs
$\a,\b\in\UP$ whose value $d(\a,\b)$ is a measure of the extent to
which the history-propositions $\a$ and $\b$ are `mutually
incompatible'.  The pair $(\UP,\D)$ is to be regarded as the
generalised-history analogue of the pair $(\L,\S)$ in standard quantum
theory.

 \item{The set $\UP$ of history-propositions is equipped with the
following logical-type, algebraic operations:
    \be
    \item{A partial order $\leq\,$. If $\a\leq\b$ then $\b$ is said to
be {\em coarser\/} than $\a$, or a {\em coarse-graining\/} of $\a$;
equivalently, $\a$ is {\em finer\/} than $\b$, or a {\em
fine-graining\/} of $\b$ . The heuristic meaning of this relation is
that $\a$ provides a more precise affirmation of `the way the universe
is' (in a transtemporal sense) than does $\b$.

    The set $\UP$ possesses a {\em unit\/} history-proposition $1$
(heuristically, the proposition about possible histories/universes
that is always true) and a {\em null history-proposition\/} $0$
(heuristically, the proposition that is always false). For all
$\a\in\UP$ we have $0\leq\a\leq 1$.
        }

    \item{There is a notion of two history-propositions $\a,\b$ being
{\em disjoint\/}, written $\a\perp\b$. Heuristically, if $\a\perp\b$
then if either $\a$ or $\b$  is `realised' the other certainly cannot be.

    Two disjoint history-propositions $\a,\b$ can be combined to form a
new proposition $\a\oplus\b$ which, heuristically, is the
proposition `$\a$ {\em or\/} $\b$'. This partial binary
operation is assumed to be commutative and associative, \ie
$\a\oplus\b=\b\oplus\a$, and $\a\oplus(\b\oplus\g)=
(\a\oplus\b)\oplus\g$ whenever these expressions are meaningful.
        }

    \item There is a negation operation $\neg\a$ such that, for all
$\a\in\UP$, $\neg(\neg\a)=\a$.
    \ee

    A crucial question is how the operations $\leq$, $\oplus$ and
$\neg$ are to be related. This issue is not addressed explicitly in
the original GH papers but we shall postulate the following, minimal,
requirements:
\begin{displaymath}
	i)\ \neg\a
        {\rm\ is\ the\ unique\ element\ in\ }\UP{\rm\ such\ that\ }
        \a\perp\neg\a {\rm\ with\ } \a\oplus\neg\a=1;
\end{displaymath}
\beq
        ii)\ \a\leq\b {\rm\ if\ and\ only\ if\ there\ exists\ } \g\in\UP
        {\rm \ such\ that\ } \b=\a\oplus\g.
\eeq
 These conditions are certainly true of, for example, subsets of paths
in a configuration space $Q$ and, together with the other requirements
above, essentially say that $\UP$ is an {\em orthoalgebra\/}; for a
full definition see \cite{FGR92}. One consequence is that
\beq
        \a\perp\b {\rm\ if\ and\ only\ if\ }\a\leq\neg\b.
\eeq
 An orthoalgebra is probably the minimal useful mathematical structure
that can be placed on $\UP$, but of course that does not prohibit the
occurence of a stronger one; in particular $\UP$ could be a lattice.
However, a property that could tell in favour of adopting
the weaker structure is that it does not seem possile to define a
satisfactory tensor product for lattices whereas this {\em is\/}
possible for orthoalgebras \cite{FGR92}.
    }

\item Any decoherence functional $d:\UP\times\UP\rightarrow\mathC$
satisfies the following conditions:
    \be
    \item {\em Null triviality\/}: $d(0,\a)=0$ for all $\a$.
    \item {\em Hermiticity\/}: $d(\a,\b)=d(\b,\a)^*$ for all
          $\a,\b$.
    \item {\em Positivity\/}: $d(\a,\a)\ge0$ for all $\a$.
    \item {\em Additivity\/}: if $\a\perp\b$ then, for
          all $\g$, $d(\a\oplus\b,\g)=d(\a,\g)+d(\b,\g)$. If
          appropriate, this can be extended to countable sums.
    \item {\em Normalisation\/}: $d(1,1)=1$.
    \ee
\ee
\noindent
In addition to the above we adopt the following GH-definition:
A set of  history-propositions $\a^1,\a^2,\ldots,\a^N$ is said to be
{\em exclusive\/} if $\a^i\perp\a^j$ for all $i,j=1,2,\ldots,N$. The
set is {\em exhaustive\/} (or {\em complete\/}) if it is exclusive and
if $\a^1\oplus\a^2\oplus\ldots\oplus\a^N=1$.

    It must be emphasised that, within this scheme, only {\em
consistent\/} sets of histories are graced with an immediate physical
interpretation. A complete set $\cal C$ of history-propositions is
said to be (strongly) consistent with respect to a particular
decoherence functional $d$ if $d(\a,\b)=0$ for all $\a,\b\in{\cal C}$
such that $\a\ne\b$. Under these circumstances $d(\a,\a)$ is regarded
as the {\em probability\/} that the history proposition $\a$ is true.
The GH-axioms then guarantee that the usual Kolmogoroff probability
sum rules will be satisfied. However, there is currently much debate
about the precise meaning in the history formalism of words like
`realised', `true', `probability' {\em etc\/}, and the precise,
contextual ontological status of a history proposition remains
elusive. This issue is of fundamental importance, but we shall not
enter the lists in the present paper.

    Our rules differ from the original GH set in several respects.
Firstly, Gell-Mann and Hartle make considerable use of the idea of
`fine-grained' histories. In our language, a fine-grained history is
an atom in the orthoalgebra $\UP$, but we have not invoked this
concept in a fundamental way since it is not clear {\em a priori\/}
that the algebras we might wish to consider are all necessarily
atomic. Even in the history version of standard quantum theory this is
problematic if one concentrates on the results of observables whose
spectra are all continuous.

    A more subtle difference between the two schemes is reflected by
our emphasis on {\em propositions\/} about histories, rather than
histories themselves. This distinction may seem pedantic but it is
appropriate when discussing inhomogeneous histories. For example, a
two-time history proposition $(\a_{t_1},\a_{t_2})$ is the sequential
conjunction ``$\a_{t_1}$ is true at time $t_1$, {\em and then\/}
$\a_{t_2}$ is true at time $t_2$'' which corresponds directly to what
one might want to call a history itself. On the other hand, if $\a$
and $\b$ are two disjoint homogeneous histories, the inhomogeneous
history proposition $\a\oplus\b$ affirms that ``either history $\a$ is
realised (\ie the sequential conjunction $\a$ is true), or history
$\b$ is realised'' and, generally speaking, this is {\em not\/} itself
a sequential conjunction but rather a proposition about the pair
$\a,\b$ of such. In other words, we feel that an `actual
universe/history' corresponds to a sequential conjunction, or analogue
thereof; all other `histories' are best viewed as propositions about
the former.

    We hope that the presentation above shows how naturally ideas of
quantum logic are suggested by the GH-scheme. Specifically, the former
uses the pair $(\L,\S)$ of single-time propositions and states; the
latter uses the pair $(\UP,\D)$ of history-propositions and
decoherence functionals. The sets $\L$ and $\UP$ are both
orthoalgebras, and the axioms for states in $\S$ and decoherence
functionals in $\D$ possess striking similarities.

    Of course, the notions of time and dynamics arise in a quite
different way within the two frameworks. In standard quantum logic,
dynamical evolution appears as a one-parameter family of automorphisms
of $\L$ whereas, in the history scheme---and in so far as the concept
is applicable at all---dynamical evolution (and the associated notion
of an `initial' state) is coded in the properties of a decoherence
functional.

    Finally, we note several key questions that need to be addressed:
\be
 \item In some instances of the general scheme there may exist a
subset $\U$ of $\UP$ that plays a role analogous to that of the
homogeneous histories (\ie sequential conjunctions) in standard
quantum theory. It is tempting to identify elements of $\U$ as
`possible universes' and to expect that all elements of $\UP$ can be
generated from $\U$ by  logical operations. The case of standard
Hamiltonian quantum theory certainly suggests that the $\oplus$ and
$\neg$ operations will generally map elements of $\U$ out of $\U$ and
into the full space $\UP$. Whether or not such a preferred subset $\U$
exists is closely connected with the question of whether there is any
{\em quasi-temporal\/} structure with respect to which an analogue of
the notion of sequential conjunction can be defined. We shall return
to this issue in section \ref{SSec:QTS}.

\item In normal quantum logic, properties like orthomodularity or
atomicity are closely related to specific physical assumptions about
relationships between states and observables/propositions. For
example, the seemingly innocuous relation $P\leq R$ implies $\s(P)\leq
\s(R)$ for all states $\s\in\S$ has significant ramifications. An
important question for the GH-scheme is whether the rules above can,
or should, be strengthened by appending additional relations of this
type. For example, should we add the requirement that $\a\leq \b$
implies $d(\a,\a)\leq d(\b,\b)$ for all $d\in\D$? This issue will be
addressed in section \ref{SSec:CE}.

\item An important challenge is to classify the set of decoherence
functionals for a given history-algebra $\UP$. For example, if the
algebra $\UP$ for some system is known to be the projection lattice
${\cal P}(\H)$ of a Hilbert space $\H$, what is the associated space
$\D$ of all decoherence functionals? In standard quantum logic,
Gleason's famous theorem \cite{Gle57} asserts that (provided
$\dim\H>2$) the set of states on ${\cal P}(\H)$ is exhausted by the
set of density matrices. Is there an analogous result for decoherence
functionals, or is the situation more $\H$-dependent than it is in
standard quantum logic? Clearly, the problem of classifying
decoherence functionals will depend closely on what extra rules might
be appended to the minimal set given above.

\item  In a standard path-integral quantum theory on a configuration
space $Q$, certain topological properties of $Q$ are reflected in the
decomposition of the path integral into sectors corresponding to
different homotopy classes for the paths \cite{DWM69,Sch81}. This
leads to an associated decomposition of a decoherence functional like
\eq{Def:dPI}. Can some general features be extracted from this special
case that could play the analogue of topological properties  for the
general scheme? A discussion of this problem is given in section
\ref{SSec:TopEff}.
 \ee

    Finally, we must ask of the extent to which it really is possible
to use a temporal form of quantum logic to put the suggestions above
on a sound footing. In particular, does the scheme work in standard
quantum theory? A key step there is to clarify the meaning of an
inhomogeneous history, and to show that the collection of homogeneous
and inhomogeneous histories can indeed be fitted together to form an
orthoalgebra, or even perhaps a lattice. The next section is devoted
to this task.

\section{A Lattice Structure for History Propositions}
\subsection{Temporal logic and tensor products in standard quantum theory}
An important step in justifying our version of the GH scheme is to
show that the set of all history propositions in standard quantum
theory can indeed be given the structure of a lattice. As this is
discussed in detail in \cite{Ish94a} we will here summarise only the
essential ideas with the aid of two-time histories.

    As we have seen, one aspect of the problem is to justify the
definitions $\C_{\a\oplus\b}:= \C_\a+ \C_\b$ and $\C_{\neg\a}:=1-\C_\a$
which, although reminiscent of quantum logic operations on
single-time projectors ($\op{P\oplus Q}=\op P+\op Q$ and $\op{\neg
P}=1-\op P$) cannot be justified as such since a product of projection
operators like $\C_\a$ is generally not itself a projector.

    To illustrate how temporal logic arises naturally consider a
two-time history $\a:=(\a_1,\a_2)$ and define $D_\a:=\a_1\a_2$. This
differs from $\C_\a$ in that it does not involve the evolution operator
$U(t_1,t_2)$ (since the operators $\a_1$ and $\a_2$ are in the
Schr\"odinger picture) and hence depends only on the propositions
themselves.

    Then $D_\a^\dagger=\a_2\a_1$ and hence $D_\a$ is not
a projector unless $\a_1$ and $\a_2$ commute. If,
nevertheless, we define $D_{\neg\a}:=1-D_\a$ we get
 \beqa
    D_{\neg\a}&=&1-\a_1\a_2\equiv(1-\a_1)\a_2+
        \a_1(1-\a_2)+(1-\a_1)(1-\a_2)       \nonumber\\
       &=&\neg\a_1\a_2+\a_1\neg\a_2+\neg\a_1\neg\a_2.
                                                        \label{Dnega}
\eeqa
On the other hand, consider a temporal-logic sequential
conjunction $A\sqcap B$ to be read as ``$A$ is true {\em and then\/}
$B$ is true''. Then $A\sqcap B$ is false if (i) $A$ is false and
then $B$ is true, or (ii) $A$ is true and then $B$ is false, or (iii)
$A$ is false and then $B$ is false; symbolically:
\beq
    \neg(A\sqcap B)=\neg A\sqcap B\ or\ A\sqcap\neg B\
                or\ \neg A\sqcap\neg B.                 \label{negAthenB}
\eeq
 This equation is remarkably similar to \eq{Dnega} and suggests that
there is indeed some close connection between temporal logic and the
general assignment $D_{\neg\a}:=1-D_\a$ even though the latter was only
justified {\em a priori\/} by reference to integration over paths in a
configuration space $Q$.

    Next we observe that something resembling \eq{negAthenB} also
appears in the quantum theory of a pair of spin-half particles. More
precisely, let $\ket\uparrow$ and $\ket\downarrow$ denote the spin-up
and spin-down states for a single particle, and let
$\ket\uparrow\ket\uparrow$ be the state of a pair of particles, both
of which are spin-up. Thus $\ket\uparrow\ket\uparrow$ belongs to the
tensor product $\H\otimes\H$ of two copies of the Hilbert space $\H$
of a single spin-half particle.  Then a pair of particles which is not
in this state is represented by
 \beq
    \neg(\ket\uparrow\ket\uparrow)=\ket\downarrow\ket\uparrow\ or\
    \ket\uparrow\ket\downarrow\ or \ket\downarrow\ket\downarrow
                                        \label{negUpUp}
\eeq
 whose appearance is suggestively similar to that of \eq{negAthenB}.
Of course, \eq{negUpUp} is not really a meaningful quantum theory
statement: at the very least, one would probably say that a state that
is not $\ket\uparrow\ket\uparrow$ is some linear combination of the
states on the right hand side of \eq{negUpUp}. Nevertheless, the
similarity between \eq{Dnega}, \eq{negAthenB}, and \eq{negUpUp} is
sufficient to suggest the key idea that, as far as its logical
structure is concerned, a two-time homogeneous history
$(\a_1,\a_2)$ can be realised by the operator
$\a_1\otimes\a_2$ on the tensor product $\H\otimes\H$ of two
copies of the Hilbert space $\H$ of the original quantum theory.

    In fact, this assignment does everything that is needed. For
example:
\bi
 \item Unlike the simple product $\a_1\a_2$, the tensor product
$\a_1\otimes\a_2$ is {\em always\/} a projection operator since
the product of homogeneous operators on $\H\otimes\H$  is defined as
$({A}\otimes{B})({C}\otimes {D}):=
{A}{C}\otimes{B}{D}$, while the adjoint operation is
$({A}\otimes{B})^\dagger:={A}^\dagger\otimes{B}^\dagger$.

\item Since $\a_1\otimes\a_2$ is a genuine projection operator
it is now perfectly correct to write $\neg(\a_1\otimes\a_2)=
1-\a_1\otimes\a_2$ on $\H\otimes\H$, and so
 \beqa
    \neg(\a_1\otimes\a_2)&=&1-\a_1\otimes\a_2		\nonumber\\
        &=&(1-\a_1)\otimes\a_2+\a_1\otimes(1-\a_2)+
            (1-\a_1)\otimes(1-\a_2)                   \nonumber \\
    	&=&\neg\a_1\otimes\a_2+
        \a_1\otimes\neg\a_2+\neg\a_1\otimes\neg\a_2
 \eeqa
which looks very much indeed like the equation \eq{Dnega} we are
trying to justify.

\item Let $(\a_1,\a_2)$ and $(\b_1,\b_2)$ be a pair of homogeneous
histories that are disjoint, \ie $\a_1\b_1=0$, or
$\a_2\b_2=0$, or both. Then, on $\H\otimes\H$, we have
$(\a_1\otimes\a_2)(\b_1\otimes\b_2)=0$ and so the projection operators
$\a_1\otimes\a_2$ and $\b_1\otimes\b_2$ are disjoint. This gives rise
to the {\em bona fide\/} equation
 \beq
    (\a_1\otimes\a_2)\oplus(\b_1\otimes\b_2) =
        \a_1\otimes\a_2 +\b_1\otimes\b_2
\eeq
which looks strikingly like the other equation $\C_{\a\oplus\b}=
\C_\a+ \C_\b$ we wish to justify.
\ei

    In summary, the correct logical structure of two-time propositions
is obtained if they are represented on the tensor-product space
$\H\otimes\H$. In particular, a homogeneous history $(\a_1,\a_2)$ is
represented by a homogeneous projector $\a_1\otimes\a_2$ and
corresponds to a sequential conjunction of single-time propositions.
However,  other projectors exist on a tensor product space that are
{\em not\/} of this simple form: specifically, inhomogeneous
projectors like $\a_1\otimes\a_2+\b_1\otimes\b_2$. These new
projectors serve to represent what we earlier called  `inhomogeneous'
history propositions.

    Note that the class operator $\C$ can now be viewed as a genuine
map from the projectors on $\H\otimes\H$ (regarded as a subset of all
linear operators on $\H\otimes\H$) to the operators on $\H$ that is
defined on homogeneous projectors by
 \beq
     \C(\a_1\otimes\a_2):= U(t_0,t_1)\a_1 U(t_1,t_2)
                        \a_2 U(t_2,t_0)
\eeq
 and then extended by linearity to the set of all projectors. This map
satisfies the relations $\C(\a\oplus\b)= \C(\a)+ \C(\b)$ and
$\C(\neg\a)=1-\C(\a)$ as genuine {\em equations\/}: it is no longer
necessary to {\em postulate\/} them via an invocation of the
analogous situation in a path-integral quantum theory.

    It is straightforward to extend the discussion above to arbitary
$n$-time histories so that a homogeneous history proposition $\h{\a}$
is represented by the projector
$\a_{t_1}\otimes\a_{t_2}\otimes\cdots\otimes\a_{t_n}$ on the tensor
product $\H_{t_1}\otimes\H_{t_2}\otimes\cdots\otimes\H_{t_n}$ of $n$ copies
of the original Hilbert space $\H$. This needs to be done for all
possible choices of temporal support, and then the whole is glued
together (see \cite{Ish94a} for details) to give an infinite tensor
product  $\otimes^\infty\H$ whose projectors carry an accurate
realisation of the space $\UP$ of all history propositions for this
Hamiltonian quantum system. In particular, the space $\UP$ is thereby
equipped with the structure of a non-distributive lattice.

    To avoid confusion it should be emphasised that, according to the
discussion above, a standard quantum-mechanical system is equipped
with two quite different orthoalgebras. The first is the normal
quantum-logic lattice $\L$ of single-time propositions associated with
the projection lattice ${\cal P}(\H)$ of the Hilbert space $\H$ of the
quantum theory, and where probabilities are given by states
$\s:\L\rightarrow [0,1]\subset\mathR$. The second orthoalgebra is the
lattice $\UP$ of `multi-time' propositions that is associated with the
projection lattice ${\cal P}(\otimes^\infty\H)$ of the (much `larger')
Hilbert space $\otimes^\infty\H$ and where probabilities are derivable
from the values of decoherence functionals
$d:\UP\otimes\UP\rightarrow\mathC$ for consistent histories. It is
important not to confuse these two algebraic structures.

    Several constructions commonly used in the consistent histories
formalism become rather natural in the language above. For example, a
complete set of exhaustive and exclusive histories corresponds simply
to a resolution of the identity operator in the space
$\otimes^\infty\H$. In particular, this accommodates rather nicely the
idea of `branch dependence'.

    To see this, consider a two-time history system represented on
$\H_1\otimes\H_2$ in which a pair $\{P_1, P_2\}$ of disjoint
propositions are the only questions that can be asked
at time $t_1$ (so that $\op P_1+\op P_2=1_{\H_1})$. Suppose we decide
that if proposition $P_1$ is found to be true at time $t_1$ then at
time $t_2$ we will test an exhaustive set of disjoint propositions
$\{Q_1,Q_2,Q_3\}$ (so that $\op Q_1+\op Q_2+\op Q_3=1_{\H_2}$),
whereas if $P_2$ is realised we will test a set $\{R_1,R_2\}$ with $
\op R_1+ \op R_2=1_{\H_2}$. Then, in this simple example of branch
dependence, the five possible homogeneous histories are the sequential
conjunctions $P_1\sqcap Q_1$, $P_1\sqcap Q_2$, $P_1\sqcap Q_3$,
$P_2\sqcap R_1$ and $P_2\sqcap R_2$, which are represented
respectively by the projection operators $\op P_1\otimes \op Q_1$,
$\op P_1\otimes\op Q_2$, $\op P_1\otimes\op Q_3$, $\op P_2\otimes\op
R_1$ and $\op P_2\otimes\op R_2$ on the Hilbert space $\H_1\otimes\H_2$.
This set of projectors is pairwise disjoint and is indeed a resolution
of the identity since
 \beq
    \op P_1\otimes\op Q_1+\op P_1\otimes\op Q_2+\op P_1\otimes\op Q_3
        +\op P_2\otimes\op R_1+ \op P_2\otimes\op R_2=1_{\H_1\otimes\H_2}.
\eeq

    It should be noted that the idea of `spreading out' time points by
using a tensor product was suggested earlier by Finkelstein in the
context of his plexor theory \cite{Fin72b,Fin78a}. Using his notation,
a homogeneous history $\a$ would be represented by the linear graph in
Figure~1 in which the vertex $t_i$ is to be associated with the
projection operator $\a_{t_i}$. In computing the decoherence
functional one associates the unitary operator $U(t_i,t_{i+1})$
with the link that joins vertex $t_{i}$ to vertex $t_{i+1}$. In the context
of his own study of quantum processes Finkelstein suggested
generalising graphs of this type to include ones with branches as in
Figure~2, and the same might be done here. For example, Figure~2
corresponds to a situation in which two systems, each with its own
internal measure of time, interact at the vertex and thereafter become
a single evolving system: a model perhaps for the
interaction of two relativistic particles.

    To form a decoherence functional using graphs like Figure~2
involves joining the free ends with an appropriate trace-type
operation, and also using a vertex-operator to link together the three
linear chains that meet. Another example of this type would be a
consistent histories version of the spin-network ideas proposed by
Penrose some years ago \cite{Pen71,Pen72,HW79}. In this case the
vertex-operator is constructed using an appropriate Clebsch-Gordon
coefficient.

    Various generalisations of the ideas above are possible. For
example, a role might be found for self-adjoint operators on
$\otimes^\infty\H$ other than projectors: presumably they represent
`multi-time' observables of some sort. Also, it is not clear that
every projector is generated by $\oplus$ and $\neg$ operations on the
homogeneous projectors; projectors that are not of this type represent
generalised history-propositions whose structure and significance
remain to be elucidated. It should also be possible to construct an
analogous scheme using a {\em continuous\/} tensor product. This could
carry the logical structure of propositions involving the values of
`time-smeared' observables like $\op q(f):=\int dt\,\op q(t)f(t)$.

\subsection{Quasi-temporal structure}
\label{SSec:QTS}
Our discussion has shown how the space $\UP$ of all history
propositions in standard quantum theory can be given the
structure of a non-distributive lattice. This justifies the postulate
that, in any generalised history theory, the space $\UP$ is at least
an orthoalgebra, if not a full lattice.

    The proof made heavy use of the existence of an ordering time
variable and, of course, that is precisely what we do not expect to be
present in a full theory of quantum gravity. However, the notions of
`time' and `homogeneous history as a temporal sequence' can be
generalised considerably whilst maintaining the ability to construct
lattice models for $\UP$, and structures of this type could appear in
quantum gravity in a variety of ways. This generalisation was
discussed in detail in \cite{Ish94a} and here we shall sketch only the
main ideas.

    The most characteristic property of a standard sequential
conjunction (\ie a homogeneous history) is the possibility of dividing
it into two such, one of which `follows' the other. Conversely, if one
sequential conjunction $\b$ follows another  $\a$, they can be
combined to give a new sequential conjunction $\a\circ\b$ that can be
read as ``$\a$ {\em and then\/} $\b$''. More precisely, the
time-ordered sequence of non-trivial propositions
$\b:=(\b_{t_1'},\b_{t_2'}\ldots\b_{t_m'})$ is said to {\em follow\/}
$\a:=(\a_{t_1},\a_{t_2}\ldots\a_{t_n})$ if $t_n<t_1'$, in which case
we define the combined sequence $\a\circ\b$ as
 \beq
    \a\circ\b:=(\a_{t_1},\a_{t_2},\ldots,\a_{t_n},
              \b_{t'_1},\b_{t'_2},\ldots,\b_{t'_m}).
\eeq
 This operation satisfies the associative law $\a\circ(\b\circ\g)
=(\a\circ\b)\circ\g$ whenever both sides are defined. Thus the space
$\U$ of homogeneous history propositions is a partial semi-group with
respect to the combination law $\circ$.

    It is clear that the temporal properties of a homogeneous history
proposition $\a\in\U$ are encoded in its temporal support. The set
$Sup$ of all temporal supports is itself a partial semi-group with
respect to the composition
 \beq
    s_1\circ s_2:=\{t_1,t_2,\ldots,t_n,t'_1,t'_2,\ldots,t'_m\}.
\eeq
 which is defined whenever the support $s_2:= \{t'_1,t'_2,\ldots,
t'_m\}$  {\em follows\/} the support $s_1:=\{t_1,t_2,\ldots,t_n\}$,
\ie when $t_n<t'_1$. The relation between the structures on $\U$ and
$Sup$ is captured by the support map $\nu:\U\rightarrow Sup$ that
associates with each homogeneous history its temporal
support. This map is a
{\em homomorphism\/} between these two partial semi-groups, \ie
$\nu(\a\circ\b)=\nu(\a)\circ\nu(\b)$ whenever both sides of the equation
are defined (in this operation, the null and unit histories are
assigned to a `base point' $*$ in $Sup$ which satisfies
$s\circ*=*\circ s=s$ for all $s\in Sup$).

    Note that, {\em ab initio\/}, a temporal support is assigned only
to homogeneous histories. Generally speaking the concept, is not
useful outside this class; for example if $\a$ and $\b$ are disjoint
homogeneous histories with different temporal supports then the
inhomogeneous history proposition $\a\oplus\b$ means that either $\a$
is realised, or $\b$ is realised, which refers to two different sets
of time points.

    This way of looking at the role of time in standard quantum theory
can be generalised considerably to give a `quasi-temporal' theory in
which the basic ingredients are (i) a partial semigroup $\U$ of {\em
possible universes\/} (a generalisation of the idea of a sequential
conjunction); (ii) a partial semigroup $Sup$ of {\em temporal
supports\/} (a generalisation of the space of finite ordered subsets
of the timeline $\mathR$); (iii) a homomorphism $\nu:\U\rightarrow
Sup$; and (iv) an orthoalgebra (possibly a full lattice) $\UP$ of
`propositions concerning possible universes' that is generated from
the preferred subset $\U$ by the logical operations $\oplus$ and
$\neg$.

    If $\a,\b\in\U$ are such that they can be combined to form
$\a\circ\b$ we say that $\a$ {\em preceeds\/} $\b$ (written as
$\a\lhd\b$), or $\b$ {\em follows\/} $\a$. Broadly speaking, the idea
is that if $\b$ follows $\a$ then there is a possibility of some
causal influence (perhaps in a rather unusual sense) of  events
`localised' in the history $\a$ on those localised in $\b$.

    It should be emphasised that the relation of `following' may not
be transitive; \ie $\a\lhd\b$ and $\b\lhd\g$ need not imply
$\a\lhd\g$; in particular, we do not suppose that $\lhd$ is a partial
order on $\U$. This is because the set of events `localised' in $\b$
that are causally affected by events in $\a$ need not include any
events that causally affect events in $\g$; an explicit example of
this phenomenon is the causal relations between compact regions in a
Lorentzian spacetime \cite{Sor93a}. We might also allow the
possibility of chains $\a^1\lhd\a^2\lhd\cdots\lhd\a^n\lhd\a^1$,
representing a generalisation of the idea of a closed time-like loop.

    In standard quantum theory, any homogeneous history $\a:=\h{\a}$
with support $\{t_1,t_2,\ldots,t_n\}$ can be written as the
composition
 \beq
    \a=\a_{t_1}\circ\a_{t_2}\circ\ldots\circ\a_{t_n}\label{Decomp_a}
\eeq
in which each single-time proposition $\a_{t_i}$, $i=1,2,\ldots,n$
is regarded as a homogeneous history whose temporal support is the
singleton set $\{t_i\}$.

    In a general history theory it is important to know when a
homogeneous history proposition can be decomposed into the product of
others. In this context we say that $\a\in\U$ is a {\em nuclear\/}
proposition if it cannot be written in the form $\a=\a_1\circ\a_2$
with both constituents $\a_1,\a_2\in\U$ being different from the unit
history $1$. Similarly, a support $s\in Sup$ is nuclear if it cannot
be written in the form $s=s_1\circ s_2$ with $s_1,s_2\in Sup$ being
different from the unit support $*$. Finally, we say that a
decomposition of $\a$ of the form $\a =\a_1\circ\a_2\circ\ldots\a_N$
is {\em irreducible\/} if the constituent history propositions $\a_i$,
$i=1\ldots N$, are all nuclear.

    Note that in the decomposition \eq{Decomp_a} of standard quantum
theory, the constituents $\a_{t_i}$ are nuclear histories in the sense
above, and $\{t_i\}$ is a nuclear support. Thus, in a general history
theory, a nuclear support is an analogue of a `point of time'; in
particular, it admits no further temporal-type subdivisions.
Similarly, a nuclear history proposition is a general analogue of a
single-time proposition. This observation raises the possibility of
finding a direct analogue of the tensor-product construction discussed
above for standard Hamiltonian physics. By this means one is enabled
to construct explicit lattice realisations for the space of history
propositions of a wide range of hypothetical systems. However note
that even if nuclear supports exist there is no {\em a priori\/}
reason why the propositions at different `points of time' should
belong to isomorphic structures, in which case one needs an obvious
generalisation of the tensor-product structure used above where the
Hilbert spaces $\H_t$ are all naturally isomorphic.

\section{The Space ${\cal D}$ of Decoherence Functionals}
\subsection{Some counter-examples in standard quantum theory}
\label{SSec:CE}
The postulated properties of decoherence functionals (see section
\ref{SSec:MinGH}) are
\be
    \item {\em Null triviality\/}: $d(0,\a)=0$ for all $\a\in\UP$.
    \item {\em Hermiticity\/}: $d(\a,\b)=d(\b,\a)^*$ for all
          $\a,\b\in\UP$.
    \item {\em Positivity\/}: $d(\a,\a)\ge0$ for all $\a\in\UP$.
    \item {\em Additivity\/}: if $\a\perp\b$ then, for
          all $\g$, $d(\a\oplus\b,\g)=d(\a,\g)+d(\b,\g)$. If
          appropriate, this can be extended to countable sums.
    \item {\em Normalisation\/}: $d(1,1)=1$.
\ee
 This is a minimal set of requirements and it is important to know if
any further conditions should be added to the list, particularly
vis-a-vis the problem of constructing and classifying decoherence
functionals.

    Of particular importance are potential relations between the
partial-order operation on $\UP$ and the ordering of the real numbers.
For example, in standard  quantum logic suppose $P,R\in\L$ satisfy
$P\leq R$. Then, since $\L$ is an orthoalgebra, there exists $V\in\L$
such that $R=P\oplus V$ and hence, if $\s\in\S$ is any state, the
additivity property $\s(P\oplus V)=\s(P)+\s(V)$ implies at once that
$\s(P)\leq \s(R)$. This is correct physically since $P\leq R$ means
that $P$ implies $R$, and it is reasonable to interpret this as
saying that, in all states, the probability that $P$ is true is less
than the probability that $R$ is true.

    It is a salutary experience to explore analogous relations in the
history formalism, and we shall now consider several plausible-sounding
inequalities  of this type, all of which turn out to be false in
standard quantum theory!

\medskip
\noindent
{\em Posited inequality 1.}
\beq
{\rm For\ all\ } d\in\D {\rm\ and\ for\ all\ }\a,\b {\rm\ with\ }
     \a\leq\b {\rm\ we\ have\ } d(\a,\a)\leq d(\b,\b).        \label{Ineq1}
\eeq
This inequality seems reasonable, not least because it clearly {\em
is\/} true when applied to sequences of projectors onto subsets of
configuration space in a path-integral quantum theory. Since, by
assumption, $\a\leq 1$ for all $\a\in\UP$, the inequality---if
true---would also imply  $d(\a,\a)\leq1$, which again looks plausible
if one bears in mind the potential probabilistic interpretation of
$d(\a,\a)$.

    On the other hand, for any given decoherence functional $d$ and
history proposition $\a$, the real number $d(\a,\a)$ is interpretable
as a probability only if $\a$ is a member of a set that is consistent
with respect to $d$. Thus, although $\a\leq\b$ is supposed to mean
that $\a$ implies $\b$ (in the sense that the latter is a
`coarse-graining' of the former, and hence constitutes a less
restrictive specification of the `history of the universe') it is not
clear that this should necessarily imply $d(\a,\a)\leq d(\b,\b)$; this
certainly cannot be derived as a theorem from the GH rules.

    In fact, the suggested inequality \eq{Ineq1} is violated in
standard quantum theory. To see this, consider a model system with an
initial state $\rho:=\ketbra\phi\phi$ (all vectors are assumed to be
of unit length) and a two-time history $(\a_1,\a_2)$ where
$\a_1(t_1):=\ketbra\chi\chi$, $\a_2(t_2):=\ketbra\psi\psi$, and where
$\ket\psi$ is chosen so that $\braket\phi\psi$=0 and $\ket\chi$ is
defined by $\ket\chi:={1\over\sqrt2}(\ket\phi+\ket\psi)$  as illustrated
in Figure~3. Then
 \beqa
\a_2(t_2)\a_1(t_1)\rho\a_1(t_1)\a_{t_2}(t_2)&=&
        \ket\psi \braket\psi\chi \braket\chi\phi\braket\phi\chi
                       \braket\chi\psi\bra\psi      \nonumber\\
            &=&{1\over 4}\ketbra\psi\psi
\eeqa
and so the corresponding decoherence functional has the value
\beq
    d(\a,\a)={\rm tr}\big(\a_2(t_2)\a_1(t_1)\rho\a_1(t_1)\a_2(t_2)\big)
            ={1\over 4}\tr(\ketbra\psi\psi)={1\over 4}.
\eeq                                                \label{dineq1}

    Now coarse-grain $\a$ by replacing the first projector $\a_1$ with
the unit operator so that the new history is $\b:=(1,\a_2)$ (this is
really a one-time history with temporal support $\{t_2\}$).
Then, by definition, $\a\leq\b$. On the other hand, the condition
$\braket\phi\psi=0$ shows at once that $d(\b,\b)=0$, hence violating
the supposed inequality $d(\a,\a)\leq d(\b,\b)$. Physically, this
is the classic quantum-mechanical effect whereby a
plane-polarised wave can be rotated through an angle of $90^0$ by
first inserting a polarisor at $45^0$ to the plane of the wave, and
then inserting another at $45^0$ to the first!

\medskip\noindent
{\em Posited inequality 2.}
\beq
    \a\perp\b{\rm\ implies\ }
        d(\a,\a)+d(\b,\b)\leq 1{\rm\ for\ all\ }d\in\D.  \label{Ineq2}
\eeq
 The analogue of this relation in standard quantum logic is $P\perp R$
implies $\s(P)+\s(R)\leq 1$. This implies that if $P$ is true with
probability $1$ in a state $\s$ (so that $\s(P)=1$) then  $R$ is
necessarily false (\ie $\s(R)=0$), and vice-versa. This gives a very
acceptable physical meaning of what it means to say that two
propositions are disjoint. It also plays a central role in Mackey's
justification of the assuption that $\L$ is orthomodular \cite{Mac63},
and it would be nice if our assumption that $\UP$ is orthomodular
could be justified in a similar way. However, as we shall see, the
inequality \eq{Ineq2} is violated by normal quantum theory.

    To find a suitable counter-example consider a pair of two-time
histories $(\a_1,\a_2)$ and $(\b_1,\b_2)$ so that
\beq
    d(\a,\a)+d(\b,\b)=\tr(\a_2(t_2)\a_1(t_1)\rho\a_1(t_1)\a_2(t_2))+
    \tr\big(\b_2(t_2)\b_1(t_1)\rho\b_1(t_1)\b_2(t_2)\big), \label{daa+dbb}
\eeq
 and make them disjoint by setting $\b_2(t_2)=1-\a_2(t_2)$. Now make
the following additional choices: $\rho=\ketbra\phi\phi$,
$\a_2(t_2):=\ketbra\psi\psi$ and $\b_1(t_1):=\ketbra\phi\phi$, where
$\braket\psi\phi=0$ and all vectors are assumed to be of unit length.
Then
 \beqa
\b_2(t_2)\b_1(t_1)&=&\big(1-\ketbra\psi\psi\big)\ketbra\phi\phi=
                    \ketbra\phi\phi=\b_1(t_1)   \nonumber\\
\b_1(t_1)\b_2(t_2)&=&\ketbra\phi\phi\big(1-\ketbra\psi\psi\big)=
                    \ketbra\phi\phi=\b_1(t_1)
\eeqa
 so that $\b_2(t_2)\b_1(t_1)\rho\b_1(t_1)\b_2(t_2)=\ketbra\phi\phi$,
whose trace is one. Thus, to violate the posited inequality \eq{Ineq2}
it suffices to find an $\a_1$ so that the trace satisfies
$\tr\big(\a_2(t_2)\a_1(t_1)\rho\a_1(t_1)\a_2(t_2)\big)>0$. But this is
easy: use the trick employed above in finding a counter-example to
\eq{Ineq1} and select $\a_1(t_1)$ to be the projector onto the
normalised vector ${1\over\sqrt2}(\ket\phi+\ket\psi)$. Then
 \beqa
    \a_1(t_1)\rho\a_1(t_1)&=&{1\over 4}\big(\ket\phi+\ket\psi\big)
                                \big(\bra\psi+\bra\phi\big)
     \ketbra\phi\phi\big(\ket\phi+\ket\psi\big)\big(\bra\psi+\bra\phi\big)
                                            \nonumber\\
        &=&{1\over 4}\big(\ket\phi+\ket\psi\big)\big(\bra\psi+\bra\phi\big)
\eeqa
 and so $\a_2(t_2)\a_1(t_1)\rho\a_1(t_1)\a_2(t_2)
={1\over4}\ketbra\psi\psi$ whose trace is $1\over 4$. Thus
$d(\a,\a)+d(\b,\b)={5\over 4}$ which violates the suggested inequality
\eq{Ineq2}.

\medskip\noindent
{\em Posited inequality 3.}
\beq
        {\rm For\ all\ } d\in\D{\rm\ and\ all\ } \g\in\UP
                {\rm\ we\ have\ }d(\g,\g)\leq 1.   \label{Ineq3}
\eeq
    This cannot be derived from \eq{Ineq1} as we have already shown
that the latter is  violated by normal quantum theory. However
\eq{Ineq3} is much weaker than \eq{Ineq1} and therefore might be
correct even if the latter is false. The inequality \eq{Ineq3} is
certainly true for any {\em homogeneous\/} history in standard quantum
theory for any unitary-evolution type of decoherence functional
$d_{(H,\rho)}$ (see \eq{Def:d}) because the norm of any projection
operator is bounded above by $1$.

    However the inequality \eq{Ineq3} is not true for all
inhomogeneous histories. Indeed, the proof that \eq{Ineq2} is violated
employed a pair of histories $\a,\b$ that are disjoint by virtue of
the fact that $\a_2\b_2=0$. This means that $d(\a,\b)=0$, and hence
 \beq
    d(\a\oplus\b,\a\oplus\b)=d(\a,\a)+d(\b,\b).     \label{daobaob}
\eeq
 But the right hand side of \eq{daobaob} was shown to have a value of
$5\over 4$, and hence the posited inequality \eq{Ineq3} is violated.
Of course, this does not mean that $d(\g,\g)>1$ for {\em all\/}
inhomogeneous histories: using the type of history constructed in the
counter-example to inequality \eq{Ineq1} it is easy to find
inhomogeneous histories of the form $\g=\a\oplus\b$ for which
$d(\g,\g)<1$.

    We note in passing that there exist `non-standard' decoherence
functionals based on non-unitary time evolution for which \eq{Ineq2}
may be violated even for homogeneous histories; see the later
discussion concerning \eq{Def:dK}.

    The fact that the, otherwise attractive, postulated
inequalities are all violated by standard quantum theory does not
necessarily prohibit one or more of them being appended to the GH
rules for the generalised theory. Indeed, one reason for invoking this
scheme as a possible framework for quantum gravity was to suggest that
normal quantum theory emerges only in some coarse-grained sense.
If so, there is no {\em a priori\/} reason for requiring standard
Hilbert-space based quantum theory to satisfy exactly all the rules of
the generalised theory: it suffices that they are obeyed in the
appropriate limit. However, this raises so many new possibilities that,
at least in this paper, we shall adopt the cautious position of
postulating no properties for the decoherence functionals that are not
possessed exactly by normal quantum theory.

\subsection{Construction of decoherence functionals}
\label{SSec:CDF}
 With this in mind we turn now to the problem of what can be said in
general about the construction of decoherence functionals that satisfy
just the minimum requirements in the GH rules. One property that
follows immediately from these is that (like the set $\S$ of states
in normal quantum logic) the space $\D$ of decoherence functionals is
a real convex set, \ie if $d_1$ and $d_2$ are decoherence functionals
then so is $rd_1+(1-r)d_2$ for any real number $r$ such that $0\leq
r\leq 1$. However, as emphasised already, we do not even know the full
classification of decoherence functionals when $\UP$ is the lattice of
subspaces of a Hilbert space, and so our answer to the general
question is necessarily very incomplete. Nevertheless, there are a few
general remarks worth making.

    Of the five listed requirements for a decoherence functional, the
most important (and potentially therefore the hardest to achieve) is
the additivity condition $d(\a\oplus\b,\g)=d(\a,\g)+ d(\b,\g)$.
Focussing on this, one class of decoherence functionals can be
obtained in the following way.

    Let $\chi:\UP\rightarrow\mathC$ be normalised by
$|\chi(1)|^2=1$ and $\chi(0)=0$, and satisfy the additive relation
\beq
    \chi(\a\oplus\b)=\chi(\a)+\chi(\b)          \label{chia0b_C}
\eeq
 so that $\chi$ is a character for the partial semigroup associated
with the $\oplus$ operation. Then a function satisfying all the
conditions for a decoherence functional can be defined by
 \beq
        d_\chi(\a,\b):=\chi(\a)^*\chi(\b).       \label{Def:dchi}
\eeq

    For example, suppose the `histories' $\a$ are all subsets of some
space $X$ equipped with a probability measure $\mu$ \ie
$\int_Xd\mu(x)=1$. Then
 \beq
        \chi_\mu(\a):=\int_{x\in\a}d\mu(x)      \label{Def:chimu}
\eeq
 satisfies \eq{chia0b_C} by virtue of the additivity of an integral
over disjoint subsets of the space on which it is defined. The
decoherence functionals of path-integral quantum theory are
essentially of this form. This example also suggests that if
quasi-temporal structure exists in the general theory it is
appropriate to add to \eq{chia0b_C} the requirement that if $\b$
follows $\a$ then
 \beq
        \chi(\a\circ\b)=\chi(\a)\chi(\b).
\eeq
In the case of path-integral quantum theory this reflects the basic
propagator property of integrals over paths in a configuration space.

    This method of generating decoherence functionals can be
generalised substantially. Specifically, let $\cal A$ be any von
Neumann algebra of operators equipped with a complex-valued function
$\phi:{\cal A}\rightarrow\mathC$ which satisfies $\phi(AB)=\phi(BA)$ and
$\phi(A^\dagger)=\phi(A)^*$ (\ie $\phi$ is a trace on $\cal A$). Now
let $\chi:\UP\rightarrow{\cal A}$ be such that
 \beqa
    &&{\rm Additivity:\ }\chi(\a+\b)=\chi(\a)+\chi(b)
                                                  \label{opchi_add} \\
    &&{\rm Evolution:\ } \a\lhd\b{\rm\ implies\ }
           \chi(\a\circ\b)=\chi(\a)\chi(\b)      \label{opchi_evo}
\eeqa
 and with $\chi(0)=0$. Then, for each such pair $(\chi,\phi)$, a
decoherence functional can be defined by
 \beq
    d_{(\chi,\phi)}(\a,\b):=\phi\big(\chi(\a)^\dagger\chi(\b)\big)
                                                \label{Def:opchi_d}
\eeq
provided $\chi$ and $\phi$ are normalised so that
$\phi\big(\chi(1)^\dagger\chi(1)\big)=1$

    This construction includes standard quantum theory if we make the
following special choices: (i) $\cal A$ is the algebra ${\cal B}(\H)$
of all bounded operators on the Hilbert space $\H$ of the quantum
mechanical system; (ii) if $\rho$ is the initial state then
$\phi(A):=\tr(A\rho)$; (iii) the map $\chi$ is the $\C$-map of
\eq{Def:C_a}. From this perspective one sees that the use of a
non-trivial von Neumann algebra $\cal A$ corresponds to a type of
non-commutative version of the integral \eq{Def:chimu}. This general
mathematical idea has been developed in depth by Connes \cite{Con94}.

    This example of standard quantum theory can itself be
generalised in a significant way. Consider the following map on a
$n$-time homogeneous history $\a=\h{\a}$
 \beq
    \chi(\a):=K(t_0,t_1)\a_{t_1} K(t_1,t_2)\a_{t_2} K(t_2,t_3)
     \ldots K(t_{n-1},t_n)\a_{t_n}K(t_n,t_{n+1})        \label{Def:chi_K}
\eeq
 where, for the moment, $K(t_{i-1},t_i)$, $i=1,2,\ldots, n+1$, is an
arbitrary set of operators on $\H$, and $t_{n+1}>t_n$ is some fiducial
`final' time (a `final' analogue of the initial time $t_0$). Equation
\eq{Def:chi_K} defines a ${\cal B}(\H)$-valued multilinear function of
the operators on $\H_{t_1}\otimes\H_{t_2}\otimes\cdots\otimes\H_{t_n}$
and hence can be extended uniquely  to a function on the inhomogeneous
histories as well.

    As it stands, if used in a decoherence functional, this function
violates a condition that seems very desirable. Namely, let
$\a$ be a homogeneous history proposition $(\a_1,\a_2,\ldots,\a_{t_{i-1}},
\a_{t_i},\a_{t_{i+1}},\ldots\a_{t_n})$
and consider the value $d(\a,\b)$ as $\a_{t_i}$ tends towards the unit
proposition $1$. The insertion of the unit proposition at any given
time should make no difference, and hence, for any $\b$, in the limit
as $\a_{t_i}=1$, $d(\a,\b)$  should equal $d(\a',\b)$ where
$\a':=(\a_1,\a_2,\ldots\a_{t_{i-1}},\a_{t_{i+1}},\ldots,\a_{t_n})$.
However, the map defined in \eq{Def:chi_K} satisfies this requirement
only if the $K$-operators obey the evolution property
 \beq
        K(t_{i-1},t_i)K(t_i,t_{i+1})=K(t_{i-1},t_{i+1})
                                                        \label{evol_K}
\eeq
 which we shall now assume to be the case.  Then, provided the
positive operator $K(t_0,t_f)^\dagger\rho K(t_0,t_f)\ge0$ has a
non-zero trace, a decoherence functional can be defined by
 \beqa
\lefteqn{
        d_{(K,\rho)}(\a,\b):={1\over\tr\big(K(t_0,t_f)^\dagger\rho
                                            K(t_0,t_f)\big)}
        \Big\{\tr\big(K(t_n,t_f)^\dagger\a_{t_n}K(t_{n-1},t_n)^\dagger
        \ldots
	}						\nonumber\\
	&&\ldots\a_{t_2}K(t_1,t_2)^\dagger\a_{t_1}K(t_0,t_1)^\dagger
         \rho K(t_0,t_1')\b_{t_1'}K(t_1',t_2')\b_{t_2'}	\nonumber\\
    	&&\qquad \ldots K(t_{m-1}',t_m')\b_{t_m'}K(t_m',t_f)\big)\Big\}
                                                        \label{Def:dK}
\eeqa
 where $t_f$ is a choice for final time that is greater than both
$t_n$ and $t_m'$. It would also be possible to add a `final' density
matrix \cite{GH90a} at the time $t_f$ but, in effect, this
is already covered by the arbitrary nature of the final $K$-operators
$K(t_n,t_f)$ and $K(t_m',t_f)$.

    Note that the `evolution' operators $K(t,s)$ need not be unitary
for the construction above to work. Thus \eq{Def:dK} could describe a
system with non-unitary time evolution. Indeed, Hartle has used a
decoherence functional of precisely this form in his analysis of the
application of the generalised quantum theory to spacetimes with
closed timelike curves \cite{Har94a}. We note also in passing that a
construction of this type would work equally well for any
quasi-temporal theory that admits nuclear supports.

    We mentioned earlier that for a standard decoherence functional of
the type $d_{(H,\rho)}$ in \eq{Def:d} (\ie corresponding to a unitary
evolution) it is trivial to see that $d_{(H,\rho)}(\a,\a)\leq1$ for
all homogeneous histories $\a$. However, this is no longer the
case for a general $K$.  Using techniques similar to those employed in
discussing \eq{Ineq1} it is not difficult to construct examples  in
which the numerator of \eq{Def:dK} is bounded below whilst the
denominator becomes arbitrarily small. Thus, even for homogeneous
histories $\a$,  $d_{(K,\rho)}(\a,\a)$
can be unbounded for non-unitary time evolution.

    It is tempting to speculate that \eq{Def:dK} is the most
general non-trivial form of decoherence functional in standard quantum
theory. If not, it would be interesting to know what additional properties of
$d\in\D$ must added to the GH rules to ensure that it {\em is\/} so.

\subsection{The adequacy of configuration space}
 One of the striking features of standard path-integration is that
when the state space of the classical system is just a cotangent
bundle $T^*Q$, {\em all\/} the quantum theory results can be obtained
from an integral over paths that lie just in the configuration space
$Q$. This suggests that in the analogous history formalism using
time-ordered strings of projectors there must be situations in which
the values of a decoherence functional $d$ on strings of projectors
onto subsets of $Q$ determines the values of $d$ on arbitrary
homogeneous histories. The collection of homogeneous histories
constructed from projections only onto subsets of $Q$ forms a
subalgebra of $\U$ that is {\em Boolean\/}, which suggests that in the
general history scheme there may be situations in which values of a
decoherence functional can be recovered from its values on some
preferred Boolean subalgebra of $\UP$.

    We will illustrate this effect in standard quantum theory and show
the precise sense in which a Boolean subalgebra of $\U$ is sufficient.
To this end consider any maximal abelian subalgebra of the
observables in the quantum theory. The maximality condition implies
the existence of a selfadjoint operator $A$ whose spectrum is totally
non-degenerate and with the property that any member of the subalgebra can
be written as a function of $A$ (a nice discussion from a physicist's
perspective is contained in \cite{Red89}). For ease of exposition we
will suppose that the spectrum of $A$ is discrete with spectral
projectors $P_m=\ketbra{m}{m}$ so that $A=\sum_ma_m P_m$ and $\sum_m
P_m=1$.

    The analogue of a string of projectors onto subsets of $Q$ is
a homogeneous history that is a time-ordered string of the projectors
$P_n$ only. Then the results of evaluating decoherence functionals on
pairs of this sort includes all the numbers
 \beq
    G_{nm}:=\tr(X P_n U(t,t')P_m)
\eeq
 where $t<t'$ are neighbouring time points and $X$ denotes the product
of all the other operators that make up the two homogeneous histories
and the initial state $\rho$.

    Now suppose that $R$ is any projector on the Hilbert space. Then
\beq
    \tr(XR)=\sum_{n,m}X_{nm}R_{mn}                  \label{tr(XR)}
\eeq
 where $X_{nm}:=\bra{n}X\ket{m}$ and $R_{mn}:=\bra{m}R\ket{n}$. On the
other hand
 \beq
    G_{nm}=\tr\big(X\ketbra{n}{n}U(t,t')\ketbra{m}{m}\big)
                            =X_{mn}U_{nm}
\eeq
which, if $U_{nm}\neq0$, can be solved for $X_{mn}$  as
$X_{mn}=G_{nm}/U_{nm}$ (no sum over $n$ or $m$), and then
substituted into \eq{tr(XR)} to give
 \beq
    \tr(XR)=\sum_{n,m}{G_{mn}\over U_{mn}}R_{mn}.   \label{tr(XR)2}
\eeq
 Bearing in mind the original meaning of $X$, we see that \eq{tr(XR)2}
is itself a decoherence functional of two homogeneous histories but
with the projector $R$ replacing the pair $P_nU(t,t')P_m$ in the
appropriate original history. Conversely, a decoherence functional of
any pair of homogeneous histories can be constructed by following the
procedure above in reverse and inserting extra time slots and
$P_nU(t,t')P_m$ factors as required. In this way any decoherence
functional can be calculated from its values on homogeneous histories
of strings of $P_n$ projectors.

    This proves our claim that it is sufficient to know the values of
a decoherence functional on strings of projectors belonging to a
maximal abelian subalgebra. The set of all such strings form a Boolean
sublattice of $\UP$. The only proviso is that the dynamical evolution
and maximal abelian subalgebra must be such that the appropriate
matrix elements $\bra{n}U(t,s)\ket{m}$ are non-zero. In practice this
is easy to ensure. For example, the case of paths in a configuration
space $Q$ can be incorporated by generalising the situation above to
include operators with continuous spectra, in which case the object of
interest is the propagator kernel
 \beq
    U(q,t;q',t'):=\bra{q'}e^{-i(t'-t)H/\hbar}\ket{q}
\eeq
and it is easy to see that the non-vanishing condition is satisfied
for a typical non-relativistic Hamiltonian $H=p^2/2m+V(q)$.

    It is worth noting that the potential role of a preferred Boolean
subalgebra of $\UP$ can be seen from another perspective. For any
given Hamiltonian there is a natural automorphism $\tau$ of $\UP$
generated by the map of homogeneous histories defined by
 \beq
    \a_{t_1}\otimes\a_{t_2}\otimes\cdots\otimes\a_{t_n}\mapsto
\a_{t_1}(t_1)\otimes\a_{t_2}(t_2)\otimes\cdots\otimes\a_{t_n}(t_n)
                                                \label{Def:tau}
\eeq
 where $\a_{t_i}(t_i)$ are the Heisenberg picture operators associated
with the Hamiltonian. It is a curious fact that this {\em single\/}
automorphism of $\UP$ endcodes {\em all\/} the information about the
dynamical evolution at all times! In particular, if one starts with a
homogeneous history $\a$ made up of strings of projectors onto a
particular abelian subalgebra then, for reasons similar to those
discussed above, the transformed history $\tau(\a)$ will generally lie
outside this preferred subset. Thus the lattice operations between
$\a$ and $\tau(\a)$ code into the {\em algebraic\/} structure of $\UP$
important information about the {\em dynamical\/} structure of the
system. This could be an important idea for future developments of the
general scheme.

\subsection{Topological effects}
\label{SSec:TopEff}
 Potential roles for topological structure in quantum theory arise in
a variety of ways. For example, the $\theta$-angle for a system with a
non simply-connected configuration space $Q$ can be justified by
considering the definition of a path-integral on $Q$. In particular,
this  suggests \cite{DWM69,Sch81} the existence of a parametrised
family of propagator functions  $K_\chi$ of the form
 \beq
    K_\chi(q_1,t_1;q_2,t_2)=
         \sum_\g\chi(\g)K_\g(q_1,t_1;q_2,t_2) \label{K=sumKg}
\eeq
 where $\chi$ is a character of the fundamental group $\pi_1(Q)$ of $Q$,
and the `partial' propagator $K_\g(q_1,t_1;q_2,t_2)$ is constructed by
integrating over paths in $Q$ that belong only to a specific
homotopy class $\g$ (identified as an element of $\pi_1(Q)$ via, for
example, a choice of reference path from $q_1$ to $q_2$).

    The same group arises in various approaches to canonical
quantisation. For example, the quantum version of a classical theory
with a configuration space $Q$ might be specified by defining the
Hilbert space of states to be the vector space of cross-sections of
some flat vector bundle over $Q$---the flatness condition guarantees
that no potential-energy term occurs in the quantum Hamiltonian (at
least, locally in $Q$) that was not present already in the classical
theory. The group $\pi_1(Q)$ appears because flat $\mathC^n$ bundles
over $Q$  are classified by the group of homomorphisms of $\pi_1(Q)$
into $U(n)$ \cite{Mil57}. A similar structure arises in the
pre-quantisation phase of geometric quantisation in which certain
types of vector bundle are constructed over the phase space of the
classical system.

    It is important to note that global properties of a classical
system can have important quantum effects even though they are not
literally `topological'. A good example is if $Q$ is the
(contractible) space of positive-definite, $n\times n$ symmetric
matrices: a natural model for the quantisation of a metric function in
quantum gravity.  This space is trivial topologically, but important
effects arise from identifying the canonical group of the system as
the semidirect product of $\mathR^{n(n+1)}$ with $GL^+(n,\mathR)$
rather than the Heisenberg group  that is appropriate for a vector
space of the same dimension \cite{Kla70,Pil82,Ish84}.

    This example belongs to the general class of system whose
configuration space is a homogeneous space $G/H$.  Systems of this
type have been quantised canonically in a variety of ways,  all of
which agree on the existence of a number of inequivalent quantum
versions of the given classical theory (defined by its Poisson algebra
of observables) that are labelled by the irreducible representations
of $H$, and which therefore generally depend only distantly on topological
properties of $G/H$.

    Note that the effects mentioned above all arise from quantising a
{\em given\/} classical system. It is a quite different matter to ask
for analogues of global or topological effects in a quantum theory
that is contructed {\em ab initio\/} in its own right. This question
is particulary interesting in the context of the GH scheme. However,
if one starts with a general $(\UP,\D)$ pair it is difficult to know
how to proceed. The space $\D$ of decoherence functionals is contractible,
and hence rather uninteresting topologically. The situation with $\UP$
is different since a space of this type could well be equipped with a
non-trivial topology. For example, the projectors on a
finite-dimensional vector space are naturally elements of various
Grassmann manifolds $U(n+m)/U(n)\times U(m)$ which are topologically
complex. However, there is no obvious role for topological
structure of this type other than, perhaps, requiring decoherence
functionals to be {\em continuous\/} functions on $\UP$ (as we assumed
implicitly in the discussion justifying the assumption that the $K$
operators satisfy the product law \eq{evol_K}). In fact, the
group-theoretical examples cited above suggest that important global
effects may arise that not related directly to any topological
properties.

    In this paper we shall concentrate only on the easier question of
identifying an analogue in the history formalism of the topological
and global properties of conventional path-integrals. The perspective
afforded by the histories approach has the additional benefit of
allowing us to exhibit a generalisation of the $\pi_1(Q)$ effects of
conventional quantum theory to a situation in which one has an
arbitrary principle bundle $H\rightarrow P\rightarrow Q$ over $Q$; the
$\pi_1(Q)$ effects come from the special case in which this is the
familiar bundle $\pi_1(Q)\rightarrow \tilde Q\rightarrow Q$ where
$\tilde Q$ is the universal covering manifold of $Q$.

    We shall restrict our attention to systems with a quasi-temporal
structure, in which case the dynamical information is coded in the
decoherence functionals (or, at least, those constructed using a form
like \eq{Def:dK}) via the evolution operators $K$ that act on the
Hilbert space $\H$ of the single-time quantum system (\ie the Hilbert
space associated with single-time propositions in the sense of nuclear
supports) and which, for all $s_1$ and $s_2$ such that $s_1\lhd s_2$,
must satisfy the relation
 \beq
    K(s_1)K(s_2)=K(s_1\circ s_2)    \label{Ks10s2}
 \eeq  in order to be consistent with the insertion of the unit
single-time proposition. Thus $K:Sup\rightarrow {\cal B}(\H)$ is a
representation on $\H$ of the partial semigroup $Sup$ of
quasi-temporal supports (in general we might have a more sheaf-like
structure in which each `time-point' is associated with its own
Hilbert space and logic of single-time propositions; in this case $K(s)$
is to be regarded as a linear map between the Hilbert spaces
associated with the two `ends' of the temporal support $s$).

    The analogous object in a path integral can be decomposed
into a sum \eq{K=sumKg} of partial propagators $K_\g$ obtained by
integrating only over paths in the homotopy class $\g$. In particular, for
$Q\simeq S^1$ we have $\pi_1(Q)\simeq Z$, and the characters
$\chi:Z\rightarrow U(1)$ are of the form
$\chi_\theta(n):=e^{in\theta}$ where the label $\theta$ runs from $0$
to $2\pi$. In this case, the partial-propagator kernels satisfy the
relation
 \beq
        K_m(q,t;q'',t'')=\sum_{n\in Z}\int_Qdq'\,
                K_{m-n}(t,q;t',q') K_n(t',q';t'',q'')   \label{KU1conv}
\eeq
 which reflects the fact that a path from $q$ to $q''$ belonging to a
given homotopy class $m$ can be formed by joining together paths from
$q$ to $q'$ and $q'$ to $q''$ subject only to the requirement that the
{\em sum\/} of the homotopy classes of the two constituent paths is
$m$.

    The situation in which we are interested is far more general but,
nevertheless, \eq{KU1conv} suggests one way in which `topological' or
`global' effects might be recognised in any quasi-temporal
history scheme centered on propagators satisfying \eq{Ks10s2}. Namely,
the situation may arise in which operators exist that do not satisfy
\eq{Ks10s2} but rather a law that is an analogue of \eq{KU1conv} in
the form
 \beq
      K_{h'}(s_1\circ s_2)=\int_H d\mu(h)\,K_{h'h^{-1}}(s_1)
                          K_{h}(s_2)           \label{Khconv}
\eeq
 where $H$ is a group with a left-invariant measure $d\mu$. In effect,
\eq{Khconv} is the group convolution law of $H$; indeed, \eq{Khconv} can be
written succinctly as
 \beq
        K(s_1\circ s_2)=K(s_1)\star K(s_2)      \label{Kstar}
\eeq
 where the $\star$ operation is the usual convolution product defined
by $(f_1\star f_2)(h'):=\int_H d\mu(h)\,f_1(h'h^{-1})f_2(h)$ but
applied to functions whose values lie in ${\cal B}(\H)$ rather than in
the complex numbers.

    At this stage two important questions must be asked. The first is
how to get from a `$\star$' representation \eq{Kstar} of the
temporal semigroup $Sup$ to a genuine representation satisfying
\eq{Ks10s2}; the second is why such objects should arise
in the first place.

    The answer to the first question uses spectral theory on the
group $H$. This involves an analogue of a Fourier transform,
$f\mapsto\tilde f$, with the property that $(f_1\star
f_2)^{\tilde{}}=\tilde{f}_1\tilde{f}_2$. For example, suppose that
$H$ is a compact Lie group, and therefore with a countable
family $\cal R$ of irreducible unitary representations. Then the
`Fourier transform' \cite{Edw72} of $f\in L^1(G,\mathC)$ is a
function $\tilde f$ of $R\in{\cal R}$ whose value $\tilde f(R)$ is a
bounded operator on the Hilbert space $\H_R$ which carries the
representation $R$ of $H$. More precisely,
 \beq
    \tilde f(R):=\int_H d\nu(h)\, f(h)R(h)
\eeq
where $R(h)$ is the operator on $\H_R$ that
represents $h\in H$. An inverse transform also exists of the form
\beq
    f(h)=\sum_{R\in{\cal R}}\tr\big(\tilde f(R) R(h)\big).
\eeq

    We use this formalism in the following way. Suppose we are given a
family of operators $K_h(s)$, $h\in H$, on a Hilbert space $\H$ that
satisfies the convolution relation \eqs{Khconv}{Kstar}. On the Hilbert
space $\H\otimes\H_R$ we can construct the `Fourier-transformed'
operator
 \beq
    K_R(s):=\int_H d\nu(h)\, K_h(s)\otimes R(h) \label{Def:KR}
\eeq
which, as required, satisfies the product law
\beq
         K_R(s_1) K_R(s_2)=K_R(s_1\circ s_2)
\eeq
 by virtue of the general property that ``the Fourier transform of a
convolution is the product of the Fourier transforms''. Thus, we
arrive at a family $K_R$, $R\in{\cal R}$, of genuine propagator
operators that can be employed in the construction of decoherence
functionals.

    The spectral theory above can be adapted easily enough to include
well-behaved discrete groups $H$; indeed, the expansion \eq{K=sumKg}
is a particular case of \eq{Def:KR} with $H=\pi_1(Q)$ and the integral
over $H$ is a sum over the irreducible one-dimensional
representations of this group.

    This result is gratifying, but it begs the question of whether
situations ever arise in which a compact Lie group is an appropriate
choice for $H$. In fact, this is commonly the case in the canononical
quantisation of a classical system whose configuration space is a
homogeneous manifold $G/H$. In the `group-theoretical' approach to
canonical quantisation, the appropriate canonical group for this
system is identified as a semidirect product of $W$ with $G$ where $W$
is a vector space that carries an representation of $G$ with the
property that one of the $G$-orbits in $W$ is diffeomorphic to
$Q\simeq G/H$ \cite{Ish84}. In \cite{LaL91} an extensive investigation
was made of the analogue of this situation in the context of path
integral quantisation, and it was shown there that each representation
$R$ of $H$ generates a genuine propagator function $K_R$ defined
formally by
 \beq
K_R(q_1,t_1;q_2,t_2):=\int_Q Dq\,e^{iS[q]/\hbar}\big(Pe^{i\int A[q]}\big)
            \label{Def:KRPI}
\eeq
 which takes its values in the operators on the Hilbert space $\H_R$
(see also \cite{Pol87,MT93}). The key quantity in \eq{Def:KRPI} is
$\big(Pe^{i\int A[q]}\big)$ which is the holonomy of the natural
$H$-invariant connection $A$ on $G/H$ along the path $q$.

    The object defined by \eq{Def:KRPI} has some obvious properties
in common with those of the $K_R$ operator constructed above,
and that the former is indeed an example of the latter can be seen
by taking the inverse Fourier transform of \eq{Def:KRPI} in the form
 \beq
    K_h(q_1,t_1;q_2,t_2)(h):=
        \sum_{R\in{\cal R}}\tr_R\big(K_R(q_1,t_1;q_2,t_2)R(h)\big)
                                            \label{Def:KhPI}
\eeq
 which, as shown in \cite{IL94b}, is essentially the path integral
$\int Dq\,e^{iS[q]}$ where the paths are restricted to those whose
holonomy is the given element $h\in H$. This is therefore a
considerable generalisation of the $K_n$ function in \eq{K=sumKg}
whose paths are required to lie in the homotopy class $n$, \ie they
have a specific holonomy with respect to the canonical flat connection
 on the universal covering bundle of $Q$ with fibre $\pi_1(Q)$.

    The holonomy properties of the object $K_h$ defined by
\eq{Def:KhPI} show that it satisfies the convolution-algebra
\eq{Khconv}. Hence this $G/H$ quantum theory does indeed provide many
examples of the type of generalised propagator we are positing for the
general scheme. It is in this sense we say that the existence in the
general theory of objects satisfying \eq{Khconv} is an analogue of
`global' or `topological' effects in normal quantum theory.

\section{Conclusion}
 We have seen how the rules proposed by Gell-Mann and Hartle for a
generalised history theory suggest strongly the use of a form of
quasi-temporal logic in which the pair $(\UP,\D)$ of
history-propositions and decoherence functionals plays a role
analogous to that of the pair $(\L,\S)$ of single-time propositions
and states in standard quantum theory. In particular, we
used a simple two-time example to illustrate the way in which the set
$\UP$ of all propositions about histories in standard quantum theory can be
realised as projectors on a tensor product Hilbert space.

    We discussed the problem of finding extensions of the GH rules for
decoherence functionals and showed how certain plausible-sounding
inequalities are in fact false in standard quantum theory. This
problem is important in the context of finding a complete
classification scheme of decoherence functionals for any given
orthoalgebra $\UP$. The non-triviality of this task is illustrated by
the existence of `non-standard' decoherence functionals \eq{Def:dK}
associated with a non-unitary time evolution. It is important to know
if decoherence functionals that correspond to unitary evolution
possess special properties such as, for example, $d(\a,\a)\leq1$ for
all homogeneous $\a$. This is particularly significant in the context
of any quantum gravity theory from which standard Hamiltonian quantum
theory is supposed to emerge in some coarse-grained sense.

    Another problem discussed in the paper is the way that
`topological' properties might appear in a formalism in which
decoherence functionals play a basic role. We have argued that, in
a quasi-temporal situation, the natural place to see the analogues of
such effects is in the properties of the generalised propagators
$K(s)$ that give an operator realisation \eq{Ks10s2} of the semi-group
$\S$ of temporal supports.

    Many problems remain to be confronted, not the least of which is
to try to apply the formalism to genuine problems in quantum
gravity that involve the use of non-standard models for space-time
such as causal sets, discrete models for quantum topology, spin
networks, and the like. In particular, we need to understand the
structure of the space $\UP$ of all `history-propositions' in
situations of this type. A key issue here is whether or not there is
some preferred Boolean sub-algbra of $\UP$ which essentially generates
the whole theory, as we saw was the case in standard Hamiltonian
quantum theory.

\bigskip
\bigskip
\noindent
{\large\bf Acknowledgements}

\noindent
We would like to thank Jim Hartle for many helpful discussions. We
gratefully acknowledge support from the SERC and Trinity
College, Cambridge (CJI) and the Leverhulme and Newton Trusts (NL).
We are both grateful to the staff of the Isaac Newton Institute for
providing such a stimulating working environment.


\begin{thebibliography}{10}

\bibitem{GH90a}
M.~Gell-{M}ann and J.~Hartle.
\newblock Quantum mechanics in the light of quantum cosmology.
\newblock In S.~Kobayashi, H.~Ezawa, Y.~Murayama, and S.~Nomura, editors, {\em
  Proceedings of the Third International Symposium on the Foundations of
  Quantum Mechanics in the Light of New Technology}, pages 321--343. Physical
  Society of Japan, Tokyo, 1990.

\bibitem{GH90b}
M.~Gell-{M}ann and J.~Hartle.
\newblock Quantum mechanics in the light of quantum cosmology.
\newblock In W.~Zurek, editor, {\em Complexity, Entropy and the Physics of
  Information, SFI Studies in the Science of Complexity, {Vol. VIII}}, pages
  425--458. Addison-Wesley, Reading, 1990.

\bibitem{GH90c}
M.~Gell-{M}ann and J.~Hartle.
\newblock Alternative decohering histories in quantum mechanics.
\newblock In K.K. Phua and Y.~Yamaguchi, editors, {\em Proceedings of the 25th
  International Conference on High Energy Physics, Singapore, August, 2--8,
  1990}, Singapore, 1990. World Scientific.

\bibitem{Har91a}
J.~Hartle.
\newblock The quantum mechanics of cosmology.
\newblock In S.~Coleman, J.~Hartle, T.~Piran, and S.~Weinberg, editors, {\em
  Quantum Cosmology and Baby Universes}. World Scientific, Singapore, 1991.

\bibitem{Har91b}
J.~Hartle.
\newblock Spacetime grainings in nonrelativistic quantum mechanics.
\newblock {\em Phys. Rev.}, D44:3173--3195, 1991.

\bibitem{GH92}
M.~Gell-{M}ann and J.~Hartle.
\newblock Classical equations for quantum systems.
\newblock 1992.
\newblock UCSB preprint UCSBTH-91-15.

\bibitem{Har93a}
J.~Hartle.
\newblock Spacetime quantum mechanics and the quantum mechanics of spacetime.
\newblock In {\em Proceedings on the 1992 Les Houches School, Gravitation and
  Quantisation}. 1993.

\bibitem{Gri84}
R.B. Griffiths.
\newblock Consistent histories and the interpretation of quantum mechanics.
\newblock {\em J. Stat. Phys.}, 36:219--272, 1984.

\bibitem{Omn88a}
R.~Omn\`es.
\newblock Logical reformulation of quantum mechanics. {I.} {F}oundations.
\newblock {\em J. Stat. Phys.}, 53:893--932, 1988.

\bibitem{Omn88b}
R.~Omn\`es.
\newblock Logical reformulation of quantum mechanics. {II.} {I}nterferences and
  the {E}instein-{P}odolsky-{R}osen experiment.
\newblock {\em J. Stat. Phys.}, 53:933--955, 1988.

\bibitem{Omn88c}
R.~Omn\`es.
\newblock Logical reformulation of quantum mechanics. {III.} {C}lassical limit
  and irreversibility.
\newblock {\em J. Stat. Phys.}, 53:957--975, 1988.

\bibitem{Omn89}
R.~Omn\`es.
\newblock Logical reformulation of quantum mechanics. {III.} {P}rojectors in
  semiclassical physics.
\newblock {\em J. Stat. Phys.}, 57:357--382, 1989.

\bibitem{Omn90}
R.~Omn\`es.
\newblock From {H}ilbert space to common sense: {A} synthesis of recent
  progress in the interpretation of quantum mechanics.
\newblock {\em Ann. Phys. (NY)}, 201:354--447, 1990.

\bibitem{Omn92}
R.~Omn\`es.
\newblock Consistent interpretations of quantum mechanics.
\newblock {\em Rev. Mod. Phys.}, 64:339--382, 1992.

\bibitem{Ish94a}
C.~Isham.
\newblock Quantum logic and the histories approach to quantum theory.
\newblock {\em J. Math. Phys.}, 1994.

\bibitem{DeW67a}
B.S. DeWitt.
\newblock Quantum theory of gravity. {I}. {T}he canonical theory.
\newblock {\em Phys. Rev.}, 160:1113--1148, 1967.

\bibitem{DeW67b}
B.S. DeWitt.
\newblock Quantum theory of gravity. {II}. {T}he manifestly covariant theory.
\newblock {\em Phys. Rev.}, 160:1195--1238, 1967.

\bibitem{Sor91b}
R.D. Sorkin.
\newblock Spacetime and causal sets.
\newblock In J.C. D'Olivo, E.~Nahmad-Achar, M.~Rosenbaum, M.P. Ryan, L.F.
  Urrutia, and F.~Zertuche, editors, {\em Relativity and Gravitation: Classical
  and Quantum}, pages 150--173, Singapore, 1991. World Scientific.

\bibitem{Bel87}
J.S. Bell.
\newblock {\em Speakable and unspeakable in quantum mechanics}.
\newblock Cambridge University Press, Cambridge, 1987.

\bibitem{VNB36}
J.~{v}on Neumann and G.~Birkhoff.
\newblock The logic of quantum mechanics.
\newblock {\em Annals of Mathematics}, 37:823--843, 1936.

\bibitem{BC81}
E.G. Beltrametti and G.~Cassinelli.
\newblock {\em The Logic of Quantum Mechanics}.
\newblock Addison-Wesley, London, 1981.

\bibitem{FGR92}
D.J. Foulis, R.J. Greechie, and G.T. {R\"uttimann}.
\newblock Filters and supports in orthoalgebras.
\newblock {\em Int. J. Theor. Phys.}, 31:789--807, 1992.

\bibitem{Mit77}
P.~Mittelstaedt.
\newblock Time dependent propositions and quantum logic.
\newblock {\em Jour.~{P}hil.~{L}ogic}, 6:463--472, 1977.

\bibitem{Mit78}
P.~Mittelstaedt.
\newblock {\em Quantum Logic}.
\newblock D.~Reidel, Holland, 1978.

\bibitem{Sta80}
{E.-W.} Stachow.
\newblock Logical foundations of quantum mechanics.
\newblock {\em Int. J. Theor. Phys.}, 19:251--304, 1980.

\bibitem{Sta81}
{E.-W.} Stachow.
\newblock Sequential quantum logic.
\newblock In E.G. Beltrametti and B.C. {van Fraassen}, editors, {\em Current
  Issues in Quantum Logic}, pages 173--191. Plenum Press, New York, 1981.

\bibitem{Gle57}
A.M. Gleason.
\newblock Measures on the closed subspaces of a {H}ilbert space.
\newblock {\em Journal of Mathematics and Mechanics}, pages 885--893, 1957.

\bibitem{DWM69}
C.~DeWitt-Morette.
\newblock {\em Ann. Inst. Henri Poincar{\'e}}, 11:11--, 1969.

\bibitem{Sch81}
L.S. Schulman.
\newblock {\em Techniques and applications of path integration}.
\newblock Wiley, New York, 1981.

\bibitem{Fin72b}
D.~Finkelstein.
\newblock Space-time code.~{III}.
\newblock {\em Phys. Rev.}, D5:2922--2931, 1972.

\bibitem{Fin78a}
D.~Finkelstein.
\newblock Process philosophy and quantum dynamics.
\newblock In C.~Hooker, editor, {\em Physical Theory as Logico-Operational
  Structure}, pages 1--18. D.~Reidel, Dordrecht, Holland, 1978.

\bibitem{Pen71}
R.~Penrose.
\newblock Angular momentum: an approach to combinatorial spacetime.
\newblock In E.~Bastin, editor, {\em Quantum Theory and Beyond}. Cambridge
  University Press, Cambridge, 1971.

\bibitem{Pen72}
R.~Penrose.
\newblock On the nature of quantum gravity.
\newblock In J.R. Klauder, editor, {\em Magic without Magic}. Freeman, San
  Francisco, 1972.

\bibitem{HW79}
L.~Hughston and R.~Ward.
\newblock {\em Advances in Twistor theory}.
\newblock Pitman, New York, 1979.

\bibitem{Sor93a}
R.D. Sorkin.
\newblock Impossible measurements on quantum fields.
\newblock In B.S. Hu and T.A. Jacobson, editors, {\em Directions in General
  Relativity, Vol. II: a Collection of Essays in honour of Dieter Brill's
  Sixtieth Birthday}. Cambridge University Press, Cambridge, 1993.

\bibitem{Mac63}
G.W. Mackey.
\newblock {\em The Mathematical Foundations of Quantum Mechanics}.
\newblock W.A.~Benjamin, New York, 1963.

\bibitem{Con94}
A.~Connes.
\newblock {\em Non Commutative Geometry}.
\newblock Academic Press, New York, 1994.

\bibitem{Har94a}
J.B. Hartle.
\newblock Unitarity and causality in generalized quantum mechanics for
  nonchronal spacetimes.
\newblock {\em Phys. Rev.}, D49, 1994.

\bibitem{Red89}
M.~Redhead.
\newblock {\em Incompleteness, Nonlocality, and Realism}.
\newblock Clarendon Press, Oxford, 1989.

\bibitem{Mil57}
J.~Milnor.
\newblock {\em Comm. Math. Helv.}, 32:215--, 1957.

\bibitem{Kla70}
J.~Klauder.
\newblock Soluble models of guantum gravitation.
\newblock In M.S. Carmeli, S.I. Flicker, and Witten, editors, {\em Relativity}.
  Plenum, New York, 1970.

\bibitem{Pil82}
M.~Pilati.
\newblock Strong coupling quantum gravity {I}: solution in a particular gauge.
\newblock {\em Phys. Rev.}, D26:2645--2663, 1982.

\bibitem{Ish84}
C.J. Isham.
\newblock Topological and global aspects of quantum theory.
\newblock In B.S. DeWitt and R.~Stora, editors, {\em Relativity, Groups and
  Topology {II}}, pages 1062--1290. North-Holland, Amsterdam, 1984.

\bibitem{Edw72}
R.E. Edwards.
\newblock {\em Integration and Harmonic Analysis on Compact Groups}.
\newblock Cambridge University Press, Cambridge, 1994.

\bibitem{LaL91}
N.P. Landsman and N.~Linden.
\newblock The geometry of inequivalent quantizations.
\newblock {\em Nucl. Phys.}, B365:121--160, 1991.

\bibitem{Pol87}
A.M. Polyakov.
\newblock {\em Gauge Fields and Strings}.
\newblock Harwood Academic Publishers, London, 1987.

\bibitem{MT93}
D.M. Mc{M}ullan and T.~Tsutsui.
\newblock On the emergence of gauge structures and generalized spin when
  quantizing on a coset space.
\newblock 1993.
\newblock Dublin Institute of Advanced Studies preprint.

\bibitem{IL94b}
C.J. Isham and N.~Linden.
\newblock 1994.
\newblock In preparation.

\end{thebibliography}
\end{document}